\documentclass[a4paper,USenglish]{lipics-v2018}

\usepackage{apxproof}
\usepackage{bbm}
\usepackage{comment}
\usepackage{currfile}
\usepackage{environ}
\usepackage{lipsum}
\usepackage{microtype}
\usepackage{pgfplots}
\usepackage[disable]{todonotes}
\usepackage{cite} 
\usepackage{xspace}

\pgfplotsset{compat=1.14}
\usetikzlibrary{calc}
\usetikzlibrary{decorations}
\usetikzlibrary{patterns}
\tikzstyle{vertex}=[draw,circle,inner sep=0pt,minimum size=6pt]
\tikzstyle{labeled}=[fill=white,inner sep=1pt,minimum size=1.6em]

\newtheoremrep{corollary}[theorem]{\bf \sffamily Corollary}
\newtheoremrep{lemma}[theorem]{\bf \sffamily Lemma}
\renewcommand{\appendixsectionformat}[2]{Appendix}
\newcommand{\apxSymbol}{\raisebox{0.3em}{$\star$}}
\newcommand{\atA}{\ifcurrfileext{tex}{\textbf{\!\!\apxSymbol\ }}{}}

\nolinenumbers

\EventEditors{Yossi Azar, Hannah Bast, and Grzegorz Herman}
\EventNoEds{3}
\EventLongTitle{26th Annual European Symposium on Algorithms (ESA 2018)}
\EventShortTitle{ESA 2018}
\EventAcronym{ESA}
\EventYear{2018}
\EventDate{August 20--22, 2018}
\EventLocation{Helsinki, Finland}
\EventLogo{}
\SeriesVolume{112}
\ArticleNo{19}

\title{Cycles to the Rescue! Novel Constraints to Compute Maximum Planar Subgraphs Fast}
\titlerunning{Cycles to the Rescue! Novel Constraints to Compute MPS Fast}

\newcommand{\ourAffiliation}{Theoretical Computer Science, Osnabrück University, Germany}

\author{Markus Chimani}\ourAffiliation{markus.chimani@uni-osnabrueck.de}{0000-0002-4681-5550}{}
\author{Tilo Wiedera}\ourAffiliation{tilo.wiedera@uni-osnabrueck.de}{0000-0002-5923-4114}{}
\funding{Supported by the German Research Foundation (DFG) project CH 897/2-1.}
\authorrunning{M.\ Chimani and T.\ Wiedera}
\Copyright{Markus Chimani and Tilo Wiedera}

\subjclass{
 \ccsdesc[500]{Mathematics of computing~Combinatorial optimization},
 \ccsdesc[300]{Mathematics of computing~Graph theory},
 \ccsdesc[500]{Theory of computation~Linear programming}
}
\keywords{
 algorithm engineering,
 graph algorithms,
 integer linear programming,
 maximum planar subgraph
}

\let\oldsum\sum
\renewcommand{\sum}{\oldsum\nolimits}

\newcommand{\bigO}{\mathcal O}
\newcommand{\ce}[1]{\multicolumn{1}{c|}{#1}}
\newcommand{\cpp}{\texttt{C++}}
\newcommand{\girth}[1]{\ensuremath{\gamma(#1)}}
\newcommand{\subgraph}[2]{\ensuremath{#1 \subseteq #2}}
\newcommand{\tablescale}{.9}
\newcommand{\tb}[1]{\textbf{#1}}
\newcommand{\skewness}{\mathit{skew}}
\newcommand{\void}{\ensuremath\varepsilon\xspace}

\definecolor{colorbrewer1}{RGB}{228,26,28}
\definecolor{colorbrewer2}{RGB}{55,126,184}
\definecolor{colorbrewer3}{RGB}{77,175,74}
\definecolor{colorbrewer4}{RGB}{152,78,163}
\definecolor{colorbrewer5}{RGB}{255,127,0}
\definecolor{colorbrewer6}{RGB}{255,255,51}
\definecolor{colorbrewer7}{RGB}{166,86,40}
\definecolor{colorbrewer8}{RGB}{247,129,191}
\definecolor{colorbrewer9}{RGB}{100,100,100}

\tikzset{every picture/.style={mark size=1.5}}

\tikzset{styleA1/.style={colorbrewer1, thick, mark=*, mark size=2pt}}
\tikzset{styleA2/.style={colorbrewer1, thick, mark=*, mark size=2pt, mark options={fill=white}}}
\tikzset{styleB1/.style={colorbrewer2, thick, mark=triangle*, mark size=3pt}}
\tikzset{styleB2/.style={colorbrewer2, thick, mark=triangle*, mark size=3pt, mark options={fill=white}}}
\tikzset{styleC1/.style={colorbrewer5, thick, mark=square*, mark size=2pt}}
\tikzset{styleC2/.style={colorbrewer5, thick, mark=square*, mark size=2pt, mark options={fill=white}}}
\tikzset{styleD1/.style={colorbrewer3, thick, mark=diamond*, mark size=3pt}}
\tikzset{styleD2/.style={colorbrewer3, thick, mark=diamond*, mark size=3pt, mark options={fill=white}}}

\pgfkeys{
 /doplot/.is family, /doplot,
 default/.style = {
  width = \textwidth,
  height = 42mm,
  title = TITLE NOT SPECIFIED,
  bars = disable,
  axisArgs = {},
  axisType = axis,
  barWidth = .7,
  percent = disable
 },
 width/.store in = \doplotWidth,
 height/.store in = \doplotHeight,
 title/.store in = \doplotTitle,
 axisArgs/.store in = \doplotAxisArgs,
 axisType/.store in = \doplotAxisType,
 bars/.store in = \doplotBars,
 percent/.store in = \doplotPercent,
 barWidth/.store in = \doplotYBarWidth
}

\tikzset{/pgf/number format/1000 sep={}}
\pgfplotsset{percentStyle/.style={
 yticklabel={\pgfmathparse{100*\tick}\pgfmathprintnumber{\pgfmathresult}\,\%},
 yticklabel style={/pgf/number format/.cd,fixed,precision=0},}
}

\newcommand{\doplot}[4][]{\pgfkeys{/doplot, default, #1}
\begin{subfigure}{\doplotWidth}
\begin{tikzpicture}
 \path[use as bounding box] (-1.2,-.9) rectangle (12.35,3);

 \newcommand{\lex}{}
 \ifthenelse{\equal{\doplotPercent}{disable}}{}{
  \renewcommand{\lex}{percentStyle,}
 }

 \ifthenelse{\equal{\doplotBars}{disable}}{}{
  \let\oldlex\lex
  \renewcommand{\lex}{\oldlex axis y line*=left,}

  \newcommand{\dolabel}{}
  \ifthenelse{\equal{\doplotBars}{nolabel}}{}{
   \renewcommand{\dolabel}{ylabel={instances},}
  }

  \begin{axis}[
    every x tick/.style={black},
    width=\doplotWidth,
    height=\doplotHeight,
    hide x axis,
    axis y line*=right,
    \dolabel
    ymin=0,
    yticklabel style={text width=1.8em}
  ] \addplot [
    draw=none,
    ybar,
    bar width=\doplotYBarWidth,
    fill,
    fill opacity=0.2
  ] table [
    col sep=comma,
    x index=0,
    y index=1] {plots/#2.csv};
  \end{axis}
 }

 \begin{\doplotAxisType}[
   every x tick/.style={black},
   width=\doplotWidth,
   height=\doplotHeight,
   \lex
   cycle list name={alg styles},
   legend cell align=left,
   legend entries={\void,c10,c10 t0 i w0,c10 t0 i s w0},
   legend style={at={(.03,.1)},anchor=south west,fill=none,postaction={fill=white,fill opacity=.5}},
   xticklabels = {#4},
   \doplotAxisArgs
  ]
  \foreach \i in {2,3,...,#3}{
   \addplot table[col sep=comma, x index=0, y index=\i] {plots/#2.csv};
  }
 \end{\doplotAxisType}
\end{tikzpicture}
\caption{\doplotTitle}
\label{sfg:#2}
\end{subfigure}}

\newcommand{\insertProofSketchFigure}{%
\begin{figure}
	\captionsetup[subfigure]{justification=centering}
	\tikzset{every node/.style={fill=black!50!white,draw,circle,inner sep=0pt,minimum size=4pt}}
	\begin{subfigure}[t]{.3\textwidth}
		\centering
		\begin{tikzpicture}
		\path[use as bounding box] (-.8,-.65) rectangle (2.5,1.6);
		\node (w1) at (0,0) {};
		\node (w2) at (2,0) {};
		\node (w3) at (1.5,1.5) {};

		\node (0) at (1,.5) {};
		\node (1) at (.75,.2) {};
		\node (2) at (1,-.5) {};

		\node (3) at (1.5,.8) {};
		\node (4) at (2,.65) {};
		\node (5) at (2.2,.8) {};

		\node (v1) at (.5, 1.25) {};
		\node (v2) at (2, -.6) {};

		\draw[very thick] (0) -- (1) -- (2) -- (0)  (w1) -- (1) -- (w2) -- (2) -- (w1) -- (0) -- (w2);
		\draw[very thick] (3) -- (4) -- (5) -- (3)  (w2) -- (4) -- (w3) -- (5) -- (w2) -- (3) -- (w3);

		\draw (w3) -- (v1) -- (w2);
		\draw (v2) -- (w2) (v2) edge[out=0, in=0] (w3);
		\draw (w1) edge[out=-90, in=180] (v2);

		\fill[color=black!10!white] (v1.center) to[bend right=120, distance=16mm] (w1.center) to (v1.center);
		\draw (v1.center) to[bend right=120, distance=16mm] (w1.center);

		\node (xw1) at (w1) {};
		\node (xv1) at (v1) {};
		\draw (v1) -- (w1);

		\node[draw=none, fill=none] (k8) at ($.5*(v1)+.5*(w1)+(-.45,.16)$) {$K_8$};
		\end{tikzpicture}
		\caption{Pseudo-Tree, Lemma~\ref{lem:ptstrength}}
		\label{sfg:pseudo-tree-strength}
	\end{subfigure}\hfill
	\begin{subfigure}[t]{.3\textwidth}
		\centering
		\begin{tikzpicture}
		\path[use as bounding box] (-.05,-.05) rectangle (2.05,2.05);
		\coordinate (1) at (0,0);
		\coordinate (2) at (2,0);
		\coordinate (3) at (2,2);
		\coordinate (4) at (0,2);

		\coordinate (5) at (1,1.75);
		\coordinate (6) at (.25,.25);
		\coordinate (7) at (1.75,.25);

		\draw[fill=black!10!white] (1) -- (2) -- (3) -- (4) -- cycle;
		\draw[fill=white] (5) -- (6) -- (7) -- cycle;

		\foreach \i in {1,...,7}{\node (v\i) at (\i) {};}

		\node[draw=none,fill=none] (k8) at (.45,1.55) {$K_7$};

		\node (w1) at (.75,.7) {};
		\node (w2) at (1.25,.7) {};

		\draw (w1) -- (v5) -- (w2);
		\draw (v6) -- (w1) -- (w2) -- (v7);
		\end{tikzpicture}
		\caption{Cycle-Edge, Lemma~\ref{lem:cestr}}
		\label{sfg:cycle-edge-strength}
	\end{subfigure}\hfill
	\begin{subfigure}[t]{.35\textwidth}
		\centering
		\begin{tikzpicture}[scale=.3]
		\path[use as bounding box] (-3.7,-3.2) rectangle (3.7,4.2);
		\node (w) at (0,0) {};
		\node (0) at (150:3) {};
		\node (1) at (90:4) {};
		\node (v1) at (90:3) {};
		\node (2) at (90:2) {};
		\node (3) at (30:3) {};
		\node (4) at (-30:4) {};
		\node (v2) at (-30:3) {};
		\node (5) at (-30:2) {};
		\node (6) at (-90:3) {};
		\node (7) at (-150:4) {};
		\node (v3) at (-150:3) {};
		\node (8) at (-150:2) {};

		\draw[bend left] (0) edge (1) (1) edge (3) (3) edge (4) (4) edge (6) (6) edge (7) (7) edge (0);
		\draw (2) edge (3) (3) edge (5) (5) edge (6) (6) edge (8) (8) edge (0);
		\draw (1) -- (v1) -- (2) (4) -- (v2) -- (5) (7) -- (v3) -- (8);

		\fill[color=black!10!white] (v1.center) to[bend right=90, distance=29mm] (w.center) to[bend left] (v1.center);
		\draw[bend left] (v1.center) to[bend right=90, distance=29mm] (w.center) to (v1.center) (w.center) edge (v2.center) (w.center) edge (v3.center);

		\node (xw) at (0,0) {};
		\node (xv1) at (90:3) {};
		\node (xv2) at (-30:3) {};
		\node (xv3) at (-150:3) {};
		\draw (0) edge[bend right=60] (2);

		\node[draw=none,fill=none] (k8) at (128:2.1) {$K_8$};
		\end{tikzpicture}
		\caption{Two-Cycles-Path, Lemma~\ref{lem:twocpstr}}
		\label{sfg:two-cycles-path-strength}
	\end{subfigure}
	\caption{Graphs in strength proofs. Bold edges in \textbf{(a)} have large weight (or are edge bundles).}
	\label{fig:strength-input}
\end{figure}}

\begin{document}

\maketitle

\begin{abstract}
 The NP-hard \emph{Maximum Planar Subgraph} problem asks for a planar subgraph~$H$ of a given graph~$G$ such that $H$ has maximum edge cardinality.
 For more than two decades, the only known non-trivial exact algorithm was based on integer linear programming and Kuratowski's famous planarity criterion.
 We build upon this approach and present new constraint classes---together with a lifting of the polyhedron---to obtain provably stronger LP-relaxations,
 and in turn faster algorithms in practice. The new constraints take Euler's polyhedron formula as a starting point and combine it
 with considering cycles in $G$. This paper discusses both the theoretical as well as the practical sides of this strengthening.
\end{abstract}

\section{Introduction}
The NP-hard \emph{Maximum Planar Subgraph} (MPS) problem is an established question in graph theory,
 already discussed in the classical textbook by Garey and Johnson~\cite{LiuGeldmacher1979, GareyJohnson1979}.
Given a graph~$G$, we ask for a largest subset~$F \subseteq E(G)$ of edges such that $F$ induces a planar graph.
By contrast, the closely related \emph{maximal} planar subgraph problem asks for a set of edges that we cannot
 extend without violating planarity and is trivially solvable in polynomial time.
The inverse measure of MPS that counts the minimum number of edges that must be removed to obtain a planar subgraph,
 is called the \emph{skewness} of $G$ and denoted by $\skewness(G)$.

There are several reasons why this problem has received a good deal of attention:
Graph theoretically, skewness is a very natural and common measure of non-planarity (like crossing number or genus).
Algorithmically, finding a large planar subgraph is central to the planarization method~\cite{batiniTalamoTamassia1984, chimaniGutwenger2012}
 that is heavily used in graph drawing:
 one starts with a large (favorably maximum) planar subgraph
  and re-inserts the deleted edges, typically to obtain a low number of overall crossings.
 In fact, this gives an approximation of the crossing number with ratio roughly $\bigO\big(\Delta \cdot \skewness(G)\big)$~\cite{chimaniHlineny17},
  where $\Delta$ is the maximum node degree.
Furthermore,
 several graph problems become easier when the input's skewness is small or constant.
 E.g., we can compute a maximum flow in time
  $\bigO\big( \skewness(G)^3 \cdot |V(G)| \log |V(G)| \big)$~\cite{hochsteinWeihe2007}\footnote{\cite{hochsteinWeihe2007} considers the crossing number; the algorithm trivially works also for the stronger parameter~skewness.}---%
  the same runtime complexity as on planar graphs if the skewness is constant.

There are several practical heuristic approaches to tackle the problem~\cite{ChimaniKleinWiedera2016}.
However, MPS is MaxSNP-hard, i.e., there is an upper bound~${\nobreak<1}$ on the obtainable
approximation ratio unless $P=\mathit{NP}$~\cite{CalinescuFernandesFinklerKarloff1998},
and there are further limits known for specific algorithmic approaches~\cite{Schmid2017, ChimaniHedtkeWiedera2016}.
Already a spanning tree gives an approximation ratio of $1/3$, the best known
ratio is $4 / 9$~\cite{CalinescuFernandesFinklerKarloff1998}, and only recently a practical
$13/33$-approximation algorithm emerged~\cite{Schmid2017}.

Considering exact algorithms, options are scarce.
Over two decades ago, an integer linear program based on Kuratowski's characterization of planarity was
introduced in~\cite{Mutzel1994}, which remained the only non-trivial exact algorithm.
Only very recently, \cite{ChimaniHedtkeWiedera2018} showed the existence of potentially feasible alternatives
to the Kuratowski-based approach, but the former still constitutes the practically by far most efficient
(and theoretically most thoroughly explored) model. All known ILP models (including those discussed in this paper) can
also directly solve the \emph{weighted} MPS, i.e., identify the heaviest planar subgraph w.r.t.\ given edge weights.

\subparagraph{Contribution.} In this paper,
 we strengthen the Kuratowski model by introducing new constraints and supplementary variables, based on analyzing the cycles
 occurring in the solutions; see Section~\ref{sec:strongerconstraints}. In particular, we show in Section~\ref{sec:hierarchy} that starting with the
 original Kuratowski model and considering cycles of growing lengths yields a natural hierarchy of ever stronger
 LP-relaxations. In Section~\ref{sec:furtherstrengthening}, we establish additional constraint classes using our cycle variables to
 further strengthen the LP-relaxations, both theoretically and practically. We show the latter property in an
 experimental evaluation in Section~\ref{sec:experiments}.
 We defer the proofs of some lemmata to the appendix, in which case we mark the lemma with~`\apxSymbol'.\todo{Refer reader to full version?}

\section{Preliminaries}

\subparagraph*{Graph Notation.}
Our non-planar input graph is called $G$.
Generally, we consider an undirected graph~$H$, with nodes $V(H)$ and edges $E(H)$, which are cardinality-2 subsets of $V(H)$.
We use $\delta_H(v)$ to denote all edges incident to node $v$ in $H$ and define the node degree $\deg_H(v):=|\delta_H(v)|$.
If $H$ is a subgraph of $G$, we write \subgraph H G.
A (sub)graph is a \emph{cycle} if it is connected and all its nodes have degree~2.
The \emph{girth} \girth H of $H$ is the length of its smallest cycle.
%
The union of two (non-disjoint) graphs~$H_1,H_2$ is denoted by $H_1 \sqcup H_2 := (V(H_1) \cup V(H_2), E(H_1) \cup E(H_2))$.
For $W \subseteq V(H)$ and $F \subseteq E(H)$ we define node- and edge-induced subgraphs $H[W] := (W, \{e \in E(H) \mid e \subseteq W\})$ and  $H[F] := \big(\bigcup_{e \in F} e, F\big)$, respectively.
We further use $H-e:=H[E(H) \setminus \{e\}]$.

Given a planar drawing~$\mathcal D$ of some planar graph~$H$,
 the cyclic adjacency order around each node in $\mathcal D$ defines an \emph{embedding}~$\pi$ of $H$.
The disjoint regions bounded by edges in $\mathcal D$ correspond to the \emph{faces} of~$\pi$; the infinite region, bounded only on the inside, is called \emph{outer face}.
The \emph{degree}~$\deg(f)$ of any face~$f$ is the number of \emph{half-edges} (``sides'' of edges) that occur on the boundary of~$f$;
a bridge occurs twice on the same face.

\subparagraph*{Linear Programming.} A \emph{Linear Program} (LP)
is a vector $c \in \mathbb R^d$ and a set of linear inequalities (\emph{constraints}) that define a polyhedron~$P$ in $\mathbb R^d$;
we ask for an element $x \in P$ that maximizes $c^\intercal x$.
An \emph{Integer Linear Program} (ILP) additionally requires the components of $x$ to be integral. For a given problem, one can establish different ILPs, so-called \emph{models}. To solve an ILP model,
one uses branch-and-bound, where dual bounds are obtained from (fractional) solutions to the \emph{LP-relaxation},
i.e., the ILP without the integrality requirements. Clearly, strong such LP-bounds are desired. We say a model $N$ is \emph{at least as strong} as a model $M$,
if $N$'s LP-relaxation gives no worse bounds than $M$'s. We say $N$ is \emph{stronger} than $M$ if, additionally, there is an instance where $N$ gives a strictly better bound.
If, in this case, $N$ arises from $M$ by adding some constraints $C$, we say $C$ \emph{strengthen} $M$.

It is often beneficial to consider only a relevant subset of constraints in the solving process,
in particular when the class of constraints is (exponentially) large.
The procedure is referred to as \emph{separation}.
We employ it on (fractional) LP-solutions for selected constraint classes.

\subparagraph*{Kuratowski Model ($\boldsymbol\void$-Model).}
The following ILP is due to Mutzel~\cite{Mutzel1994}. Jünger and Mutzel showed
that both constraint classes below form facets of the planar subgraph polytope~\cite{JuengerMutzel1996}.
We use solution variables~$s_e \in \{0,1\}$ (for all $e \in E(G)$) that are $1$ if and only if edge~$e$ is deleted, i.e., \emph{not} in the planar subgraph.
(In \cite{Mutzel1994}, equivalent variables $x_e:=1-s_e$ are used.)
The objective minimizes the skewness---thus maximizes the planar subgraph---and is given~by
\[
 \min \sum_{e \in E(G)} w(e)\, s_e.
\]
Thereby, we may consider edge weights $w$; they are $1$ in case of the traditional unweighted MPS problem.
For a given subset~$F \subseteq E(G)$ of edges, we define $s(F) := \sum_{e \in F} s_e$ as a shorthand.
We can always use Euler's bound on the number of edges in planar graphs:\begin{equation}
 s\big(E(G)\big) \geq |E(G)| - (3n-6) + \mathbbm{1}_{G\text{ is bipartite}}(n - 2). \label{eq:euler}
\end{equation}
By Kuratowski's theorem~\cite{Kuratowski1930}, a graph is planar if and only if it neither contains a subdivision of a $K_5$ nor of a $K_{3,3}$.
Hence, it suffices to ask for any member of the (exponentially large) set~$\mathcal K(G)$ of all Kuratowski subdivisions that at least one of its edges is deleted:
\begin{align}
 s\big(E(K)\big) &\geq 1\quad & \forall K \in \mathcal K(G). \label{eq:kurat}
\end{align}
Clearly, \eqref{eq:kurat} are too many constraints to use all explicitly.
Instead, we identify a sufficient subset of constraints via a (heuristic) separation procedure:
 we round the fractional solution and obtain a graph that can be tested for planarity.
 If it is non-planar, we extract a Kuratowski subdivision.
 This method does neither guarantee to always find a violated constraint if there is any, nor that
 the identified subdivision in fact corresponds to a \emph{violated} Kuratowski constraint.
 Still, since it has these guarantees on integral solutions, it suffices to obtain an exact algorithm.
Over the years, the performance of this approach was improved by
 strong preprocessing~\cite{ChimaniGutwenger2009},
 finding \emph{multiple} Kuratowski subdivision in linear time~\cite{ChimaniMutzelSchmidt2007},
 and strong primal heuristics~\cite{ChimaniKleinWiedera2016}. We use all these identically in all considered algorithms.

The Kuratowski-model forms the basis of our extensions. As such, we denote it, without any of the below extensions,
by `$\void$-model'.

\section{Stronger Constraints Based on Cycles}\label{sec:strongerconstraints}
We now present new constraints for the planar subgraph polytope (or a lifted version thereof).
All but the first class require the introduction of new variables based on cycles, leading to the \emph{cycle model}.
For each constraint class we first give some motivation and intuition for its feasibility, before discussing its technical details.
We then describe---provided the class is large---separation routines that quickly identify violated constraints, and usually show that
 it strengthens our ILP model.

\newcommand{\lpobj}{\ensuremath{\mathrm{OBJ}}}
\begin{toappendix}
 The appendix consists mainly of proofs for the strength of certain classes of constraints.
 Such proofs always proceed in the following manner:
 First, we describe an integrally edge-weighted graph that is used as input.
 Note that (integral) edge-weights are naturally obtained by contracting parallel $2$-paths in the input
  and other preprocessing techniques.
 We restrict ourselves to instances that cannot be reduced by standard techniques~\cite{ChimaniGutwenger2009}.
 Next, we give an LP-feasible solution with objective value $\lpobj$ for the model that does not use the new constraints.
 In particular, we also show that the solution satisfies \emph{all} traditional Kuratowski constraints~\eqref{eq:kurat}.
 Finally, we show that there is no LP-feasible solution with objective value~\lpobj\ when using (a subset of) the new constraints.
\end{toappendix}

\subparagraph*{Generalized Euler Constraints.}
We know from~\cite{JuengerMutzel1996} 
that inequality~$|E(G)| \leq 2|V(G)|-4$ is facet-defining for complete biconnected graphs.
We are interested in a class of similar constraints for dense subgraphs with large girth. The following lemma is folklore:
\begin{lemma}
 A planar graph~$G$ has at most $\big(|V(G)|-2\big) \girth G / \big(\girth G-2\big)$ edges.
 \begin{proof}
  Let $n := |V(G)|$, $m := |E(G)|$, and $\pi$ denote an embedding of $G$.
  For any face of $\pi$ we require at least $\girth G$ half-edges.
  Thus, the number~$f$ of faces in $\pi$ is bounded by $f \leq 2m / \girth G$.
  Using Euler's formula, we obtain $n - m + (2m / \girth G) \leq 2$, the claimed results follows when solving for $m$.
  \end{proof}
\end{lemma}
We can thus derive a feasible \emph{generalized Euler constraint} for any subgraph $H\subseteq G$:
\begin{align}
  |E(H)| - s\big(E(H)\big) &\leq \big(|V(H)|-2\big)\girth H/\big(\girth H-2\big) & \forall H \subseteq G \label{eq:gen-euler}
\end{align}
We note that this bound can sometimes be improved:
 for constraints~\eqref{eq:gen-euler} to be satisfied with equality it is necessary that
 $V(H) \equiv 2\ (\mathrm{mod}\ \girth H - 2)$ if $\girth H$ is odd
 and $V(H) \equiv 2\ \big(\mathrm{mod}\ (\girth H -2)/2\big)$ otherwise \cite{FernandezSiegerTait2017}.
However, we did not implement this in our algorithms.
\todo{Do implement and test?}

\begin{lemmarep}\atA
 The generalized Euler constraints~\eqref{eq:gen-euler} strengthen the \void-model.
\end{lemmarep}

\begin{proofsketch}
 $K_{3,3,1}$ contains a $K_{3,4}$ that prohibits the otherwise feasible solution $3/2$.
\end{proofsketch}

\begin{proof}
 Let $G$ denote the $K_{3,3,1}$ (see Fig.~\ref{fig:gen-euler-strength}).

 Consider the \void-model on $G$. We show the fact for $\lpobj = 3 / 2$.
 Since $\girth{G} = 3$,
  the traditional Euler constraint only gives an upper bound of~$15$ on the number of edges in
  the planar subgraph, but $|E(G)| = 15$.
 Let $v$ denote the node in the cardinality-$1$ partition
  and $M$ denote a perfect matching in $G':=G - v$.
 For any cardinality-2 subset $S\subset M$, the graph $G-S$ is planar (see Fig.~\ref{sfg:k-3-3-matching}).
 Hence, we satisfy all Kuratowski constraints by setting $s_e = 1 / 2$ for all $e \in M$.
 It follows that an LP-value of $3 / 2$ is feasible without the generalized Euler constraints.

 Consider an edge cut~$F$ in $K_{3,3,1}$ that partitions the nodes $V(G')$ such that $E(G')$
 is fully contained in $F$, and places $v$ in either of the two partitions.
 $F$~induces a~$K_{3,4}$-subgraph~$H$ with $\girth{H}=4$ (see Fig.~\ref{sfg:k-3-4-subgraph}).
 The generalized Euler constraint on $H$ is $s(E(H)) \geq 2$, hence improving the dual bound.
 \begin{figure}[h]\centering 
	\begin{subfigure}[t]{.44\textwidth}\centering
		\begin{tikzpicture}[every node/.style=vertex]
		\path[use as bounding box] (-1.1,-1.5) rectangle (2.2,1.5);
		\node[fill=gray] (b1) at (30:1) {};
		\node (a2) at (90:1) {};
		\node[fill=gray] (b2) at (150:1) {};
		\node (a3) at (210:1) {};
		\node[fill=gray] (b3) at (270:1) {};
		\node (a1) at (330:1) {};
		\node[rectangle] (x) at  (0:2) {};
		\draw (a1) -- (b1) -- (a2) -- (b2) -- (a3) -- (b3) -- (a1);
		\draw[dashed] (a1) -- (b2) (b1) -- (a3) (b3) -- (a2);
		\draw (x) -- (b1) (x) edge[out=90, in=90] (b2) (x) edge[bend left] (b3);
		\draw (x) -- (a1) (x) edge[out=-90, in=-90] (a3) (x) edge[bend right] (a2);
		\end{tikzpicture}
		\caption{Edges of the maximum matching~$M$ in~$G'$ are dashed.}
		\label{sfg:k-3-3-matching}
	\end{subfigure}\hspace{.1\textwidth}%
	\begin{subfigure}[t]{.44\textwidth}\centering
		\begin{tikzpicture}[every node/.style=vertex]
		\path[use as bounding box] (-.2,-.5) rectangle (3.2,2.5);
		\node[fill=gray] (b1) at (0, 0) {};
		\node[fill=gray] (b2) at (1, .25) {};
		\node[fill=gray] (b3) at (2, .5) {};
		\node (a1) at (0, 2) {};
		\node (a2) at (1, 1.75) {};
		\node (a3) at (2, 1.5) {};
		\node[rectangle] (x) at  (3, 1.5) {};
		\draw (a1) -- (b1) -- (a2) -- (b2) -- (a3) -- (b3) -- (a1);
		\draw (a1) -- (b2) (b1) -- (a3) (b3) -- (a2);
		\draw[dashed] (x) edge[out=90,in=45] (a1) (x) edge[out=120,in=40] (a2) (x) -- (a3);
		\draw (x) edge[out=-90,in=-45] (b1) (x) edge[out=-120,in=-40] (b2) (x) -- (b3);

		\draw[thick,decorate,decoration={zigzag,segment length=4,amplitude=.9}] (-.25,1) -- (3.6,1);
		\node[fill=white,draw=none] (label) at (3.6,1) {$F$};
		\end{tikzpicture}
		\caption{The $K_{3,4}$-subgraph (dashed edges are removed from~$K_{3,3,1}$).}
		\label{sfg:k-3-4-subgraph}
	\end{subfigure}
	\caption{The input~$K_{3,3,1}$ for showing the strength of the generalized Euler constraints.}
	\label{fig:gen-euler-strength}
\end{figure}
\end{proof}

We separate constraints~\eqref{eq:gen-euler} heuristically by seeking dense, high-girth subgraphs using two different methods.
First, using the current fractional solution, we assign weight $1-s_e$ to each edge $e$ 
  and approximate a maximum cut~\cite[Section~6.3]{mitzenmacherUpfal2005},
  obtaining a girth-$4$ subgraph. If (after postprocessing, see below) this does not yield a
  violated constraint, we try a second method: We set a target girth~$\mu$ and
 iteratively add edges in ascending order of their LP-value to an initially empty graph,
  while updating the shortest paths between all node pairs.
 Upon adding an edge~$e$, we check whether $e$ would create a cycle of length $< \mu$, in which
 case we discard $e$ instead.
 We may repeat this process for different values of $\mu$.
 After each of the above attempts, we apply a postprocessing:
 Let $H$ denote a girth-$\mu$ subgraph.
 The \emph{contribution} of a node $v \in V(H)$ is defined by $|\delta_H(v)| - \sum_{e \in \delta_H(v)} s_e - \mu / (\mu-2)$.
 We iteratively remove nodes with negative contribution from $H$. In particular, this will remove all degree-$1$ nodes.

\subsection{Cycle Model}\label{sec:cycles}
 We now want to bound the number of edges in the planar subgraph by the number of its small faces.
 Even though compelling from a theoretical standpoint,
  it is infeasible to generate all potential faces of all planar subgraphs of a given graph (already for bounded length).
 However, we know that traversing the border of any face of a spanning subgraph~$H$ traverses at least one cycle if $H$ is not a tree.
 We will relate the number of small faces in any planar subgraph of a graph~$G$ to the number of small cycles in~$G$.
 One may also view this as a way to further generalize Euler constraints:
  many---in particular sparse---graphs have low girth only due to very few cycles of small length.

We may assume any (maximal) primal solution to be connected and non-outerplanar as it could be trivially improved otherwise.
 Also observe that we cannot require faces to uniquely map to cycles in general.
 Consider for example
  a cycle graph (two faces with the same cycle) or
  a non-biconnected graph (each cut-node occurs twice in at least one face; cycles contain nodes at most once)
Note that there are biconnected graphs that have no biconnected MPS.

\begin{lemma} \label{lem:cycles-faces}
For every connected, planar but non-outerplanar subgraph~$H$ of $G$,
 there exists an embedding of $H$ such that
 we can assign a unique cycle~$\alpha$ to every face~$f$ where all edges of $\alpha$ occur on the boundary of $f$.
\begin{proof}
 Let $H\subset G$ be as defined in the claim.
 There exists some biconnected component~$B^\star$ of $H$ that is neither a cycle nor an edge since $H$ is not outerplanar.
 Choose an embedding of $H$ and pick some face of $B^\star$ as the outer one.
 For every biconnected component~$B$ that is not just an edge, we iterate over the inner faces of $B$.
 Each inner face~$f$ of $B$ directly corresponds to a cycle as a biconnected graph contains neither cut-nodes nor bridges.
 (Observe that an inner face of $B$ might in fact be much larger in $H$ since we ignore other components nested in this face.)
 Ultimately, we assign the cycle induced by the outer face in $H$ to the (last remaining) outer face.
 Since $B^\star$ is not a cycle we do not assign any cycle twice.
\end{proof}
\end{lemma}
We denote the number of faces whose degree satisfies some property $\mathcal{P}$ by $f_\mathcal{P}$.
\begin{lemma}
 Given a connected, planar graph~$H$ on $n$ nodes and $m$ edges,
  for each embedding of $H$ with exactly $f_{=d}$~faces of degree~$d \in \{3,4\ldots,2m\}$,
  we have \begin{align}
    m = 3n-6 -\sum_{d=3}^{2m} (d-3) f_{=d}. \label{eq:euler-n-face-deg}
  \end{align}
 \begin{proof}
  Every face in any embedding of~$H$ has degree at least $3$ and at most $2m$.
  For every face~$f$ of degree~$d$ we can add $d-3$ edges that split $f$ into $d-2$ triangles without violating planarity.
  After performing this operation for each face we obtain a planar triangulated graph, i.e., a graph that has exactly $3n-6$ edges.
 \end{proof}
\end{lemma}

 Let $\mathcal C_d(G)$ denote all cycles of length $d$ in $G$.
 We set $D \geq 3$ to the \emph{maximum cycle length} that we want to investigate;
  this parameter will control the number of additionally generated variables.
 Let $\mathcal C_{\leq D}(G)$ denote the set of cycles with length at most $D$.
 For every cycle $\alpha \in \mathcal C_{\leq D}(G)$ we generate a variable $c_\alpha \in \{0,1\}$.\footnote{Intuitively, we want $c_\alpha=1$ if and only if $\alpha$ is (part of) a face, see below for details. In terms of correctness,
  we need not but can actively force these variables to be binary, cf.\ Section~\ref{sec:experiments}.}
 We force such a variable to $0$ if any edge of the respective cycle is removed and allow at most two cycles per edge in the MPS:
\begin{align}
 \sum_{\alpha \in \mathcal C_{\leq D}(G)\colon e \in E(\alpha)} c_\alpha &\leq 2\, (1-s_e) & \forall e \in E(G) \label{eq:at-most-2-cycles}
\end{align}
Note that constraints~\eqref{eq:at-most-2-cycles} resemble the requirement for each edge to appear in at most two faces (subject to Lemma~\ref{lem:cycles-faces}). We discuss its correctness below.
Let $c(d) := \sum_{\alpha \in \mathcal C_d(G)} c_\alpha$.
\begin{lemma} \label{lem:cycle-vars-faces}
 For every connected, planar but non-outerplanar subgraph~$H$ of~$G$,
  there exists an embedding $\pi$ of $H$ and a feasible assignment w.r.t.\ \eqref{eq:at-most-2-cycles} of cycle variables such that for each $d \leq D$ the number~$f_{=d}$
  of faces with degree $d$ in $\pi$ is bounded from above by \begin{equation*}
  f_{\leq d} \leq \sum_{k=3}^d c(k)\text{,\quad or equivalently}\quad  f_{=d} \leq \sum_{k=3}^d c(k) - f_{< d}. \label{eq:deg-d-ub}
  \end{equation*}
\begin{proof}
 We assign cycle variables following the proof of Lemma~\ref{lem:cycles-faces}.
 Hence, there is a unique cycle variable assigned to each face such that the length of the cycle is at most the degree of its face.
 The variable assignment is feasible since we pick only edges contained in $H$ and pick at most two cycles incident with any such edge.
\end{proof}
\end{lemma}

\begin{theorem}
 For any maximum planar subgraph of a graph~$G$ on $n$ nodes and $m$ edges there exists a feasible variable assignment that satisfies \eqref{eq:at-most-2-cycles} and the \emph{cycle constraint}\begin{equation}
  (D-1)\big(m-s(E(G))\big) \leq (D+1)(n-2)+\sum_{d=3}^D (D+1-d) c(d). \label{eq:ub}
 \end{equation}
 \begin{proof}
Starting with \eqref{eq:euler-n-face-deg} on any connected, planar subgraph of~$G$ that has $m-s(E(G))$ edges,
 we relax the equality by using the same coefficient for all faces of large degree as in
\begin{equation*}
 m-s(E(G)) \leq 3n-6-\sum_{d=3}^{D}(d-3) f_{=d} - (D-2) f_{>D}.
\end{equation*}
By replacing $f_{>D}$ in Euler's formula, $(f_{>D} + f_{\leq D}) + n - \big(m - s(E(G))\big) = 2$, we obtain
\begin{equation}
 (D-1)\big(m-s(E(G))\big) \leq (D+1)(n-2)+\sum_{d=3}^{D}(D+1-d) f_{=d}. \label{eq:ub-fD}
\end{equation}
The claimed cycle constraint is finally obtained by applying Lemma~\ref{lem:cycle-vars-faces} to iteratively replace $f_{=D'}$ for $D' = D,D-1,\ldots,4,3$ by the upper bound (note that $f_{<3}=0$),
 as sketched below for the (generalized, iteratively re-appearing) rightmost summand of~\eqref{eq:ub-fD}:
\begin{equation*}
 \sum_{d=3}^{D'} (D'+1-d) f_{=d} \leq \sum_{d=3}^{D'-1} \big((D'-1)+1-d\big) f_{=d} + \sum_{d=3}^{D'} c(d) \qedhere
\end{equation*}
\end{proof}
\end{theorem}

\subsection{Relaxations and $\mathbf D$-Hierarchy}\label{sec:hierarchy}
\newcommand{\CM}[1]{\ensuremath{\mathrm{CM}_{#1}}}
We now turn our attention to LP-relaxations of the cycle model.
We show that there is a hierarchy of gradually stronger LPs induced by the maximum cycle length~$D$.
Let the \emph{cycle model} $\CM D$ consist of the \void-model, the cycle variables for cycle lengths up to $D$, and the corresponding constraints \eqref{eq:at-most-2-cycles},\eqref{eq:ub}.
If $D=2$ were allowed, $\CM 2$ 
would be exactly the~\void-model.

\begin{lemma}\label{lem:weighted-sum-of-cycle-vars}
 For any solution to the relaxation of \CM D, it holds that \[
  \sum_{d=3}^D (d-2)c(d) \leq 2n-4.
 \]
 \begin{proof}
  Assume the contrary, $\sum_{d=3}^D(d-2)c(d) > 2n-4.$
  It follows that
   $\sum_{d=3}^Ddc(d) > 2n - 4 + 2\sum_{d=3}^Dc(d)$ and hence
   $m-s(E(G)) > n - 2 + \sum_{d=3}^Dc(d)$ by the sum of constraints~\eqref{eq:at-most-2-cycles}.
  Plugging this bound on the number of edges into the cycle constraint~\eqref{eq:ub},
   we obtain $\sum_{d=3}^D(d-2)c(d) < 2n-4$, a contradiction.
 \end{proof}
\end{lemma}

It is not immediately clear,
 that decreasing the maximum cycle length maintains LP-feasibility,
 as some variables are removed and the cycle constraint is replaced.
By employing Lemma~\ref{lem:weighted-sum-of-cycle-vars}, we can show the following fact.

\begin{lemmarep}\atA
 Model $\CM {D+1}$ is at least as strong as $\CM D$.
\end{lemmarep}

\begin{proof}
  Let $A := (\bar s, \bar c)$ denote a feasible variable assignment for $P'$.
  By eliminating cycle variables of length $D+1$ from $A$ we obtain a feasible variable assignment for $P$ with the same objective value:
  as we do not change the value of $\bar s$, we respect all Kuratowski constraints.
  The sum of cycle variables for each edge does not increase, thus, constraints~\eqref{eq:at-most-2-cycles} are respected as well.
  It remains to show that the cycle constraint is satisfied which is obtained in the following way, starting with the trivial identity:
  \begin{align*}
   \sum_{d=3}^{D} D(D+1-d) c(d) &= \sum_{d=3}^{D} D(D+1-d) c(d) \\
   \intertext{We apply Lemma~\ref{lem:weighted-sum-of-cycle-vars} for $D+1$.}
   2n-4 + \sum_{d=3}^D D(D+1-d) c(d) &\geq \sum_{d=3}^{D+1} (D-1)(D+2-d) c(d)\\
  \end{align*}
  Adding $(D^2+D)(n-2)$ to both sides and afterwards dividing by $D(D-1)$ gives the right-hand sides of equation~\eqref{eq:ubx}, divided by factor~$D-1$ (resp.\ $D$).
  \begin{align*}
   &\left( (D+1)(n-2) + \sum_{d=3}^D (D+1-d) c(d) \right) \cdot (D-1)^{-1}
   \\ \geq &\left( (D+2)(n-2) + \sum_{d=3}^{D+1} (D+2-d) c(d) \right) \cdot D^{-1}
  \end{align*}
  The upper bound on the number of edges due to $D$ is no less than that due to $D+1$.
\end{proof}

\begin{lemma}
 Model $\CM {D+1}$ is stronger than $\CM D$.
 \todo{What about pseudo-trees?}
 \begin{proof}
 Consider the complete graph~$K_k$ on $k \geq 5$~nodes. Pick any number $\mu \geq D+1$.
We subdivide every edge of $K_k$ using $\xi := \lfloor \mu / 3 \rfloor$ additional nodes.
The resulting graph~$K_k^\mu$ has girth at least $\mu$, i.e., it has no cycles of length $\leq D$.
We observe that $\skewness(K_n^\mu)=\skewness(K_k)=k(k-1)/2 - 3k+6$, independent of $\mu$.
\todo{This is no NPC.}
We show that increasing the maximum cycle length from $D$ to $D+1$ cuts off all previously optimal LP solution.

  Since $K_k^\mu$ has girth $\mu$ there can be at most $(|V(K_k^\mu)|-2) \mu / (\mu -2)$ edges in any planar subgraph.
  As there are no cycle variables, the cycle constraint~\eqref{eq:ub} approaches this value from above for increasing $D$.
  Any feasible solution that tightly satisfies the cycle constraint is an optimal one.
  The Kuratowski constraints~\eqref{eq:kurat} on the other hand are already satisfied by deleting each edge partially with $s_e = 1 / (9\xi)\ \forall e \in E(K_n^\mu)$,
   since each subdivision requires at least $9\xi$ edges, still allowing LP-solutions with value~$k(k-1) / 18$.
 \end{proof}
 \label{lem:inc-D-has-stronger-bound}
\end{lemma}

Overall, increasing the maximum cycle length strengthens our LP relaxations (leading to fewer LP-computations), but this comes
 at the cost of 
 increasing the variable space (leading to slower LP-computations).
It is imperative to find a good trade-off between these two.

\subsection{Strengthening the Cycle Model}\label{sec:furtherstrengthening}

We now extend the cycle model further by introducing new constraint classes. Only the first such extension requires yet additional variables.

\subparagraph*{Pseudo-Tree Extension.}
Observe that degree-$1$ nodes in the solution deteriorate the cycle constraint's bound:
given a face $f$ that contains a degree-$1$ node, we can set the variable of a cycle with length at most $\deg(f)-2$  to $1$.
We introduce new variables $t_v \in \{0,1\}$ for all $v \in V(G)$ and $t_{vw} \in \{0,1\}$ for all $v,w \in V(G)$ with $\{v,w\}\in E(G)$. They label nodes and directed edges (\emph{arcs}) as \emph{pseudo-trees}:
any node with at most one unlabeled neighbor (in particular any degree-1 node) is to be labeled.
This can be achieved by:
\begin{align}
 t_{vw} + t_{wv} &\leq 1 - s_{\{v,w\}} &&\forall \{v,w\} \in E \label{eq:pc-one-arc}\\
 \sum_{w \in N(v)}t_{vw} &\geq t_v &&\forall v \in V(G) \label{eq:pc-prop-adj}\\
 t_v + \deg_G(v) - \sum_{w \in N(v)}t_{wv} - \sum_{w \in N(v)}s_{vw} &\geq 2 &&\forall v \in V(G) \label{eq:pc-prop-node}
\end{align}
Constraints~\eqref{eq:pc-one-arc} allow at most one tree-arc for any edge and none for deleted edges.
We force tree nodes to propagate along one outgoing arc by constraints~\eqref{eq:pc-prop-adj}.
Finally, constraints~\eqref{eq:pc-prop-node} label degree-1-nodes and nodes where all (but one) neighbor is labeled.
Now, we may subtract $\sum_{v \in V(G)} t_v$ nodes (and the same number of edges) from \eqref{eq:ub} to obtain a stronger bound:
\begin{corollary}
 The \emph{extended cycle constraint}, given below, is feasible.
 \begin{align}
  (D-1)\big(m - s(E(G))\big) &\leq (D+1)(n-2) + \sum_{d=3}^D (D+1-d) c(d) -2\sum_{v \in V(G)}t_v \label{eq:ubx}
 \end{align}
\end{corollary}
Alternatively, we may use a less sophisticated approach that does not model propagation but labels only degree-$1$-nodes.
In this case, it suffices to add variables~$t_v \in \{0,1\}$, $\forall v \in V(G)$, and constraints~\eqref{eq:pc-prop-node}, assuming $\sum_{w \in N(v)}t_{wv} = 0$.


\begin{lemmarep}\atA\label{lem:ptstrength}
 The pseudo-tree extension, i.e., constraints~\eqref{eq:pc-one-arc}--\eqref{eq:ubx} together with the $t$-variables, strengthens \CM 3.
 This already holds for the approach without propagation.
\end{lemmarep}

\begin{proofsketch}
 We use the graph given in Fig.~\ref{sfg:pseudo-tree-strength}: any MPS of it has a degree-$1$ node.
\end{proofsketch}

\begin{proof}
\begin{figure}\centering
	\begin{tikzpicture}[every node/.style=vertex]
	\path[use as bounding box] (-1.55,-1.75) rectangle (9.5,2);
	\node[labeled] (w1) at (0,0) {$w_1$};
	\node[labeled] (w2) at (4,0) {$w_2$};
	\node[labeled] (w3) at (8,0) {$w_3$};

	\node (0) at (2,.5) {};
	\node (1) at (2,0) {};
	\node (2) at (2,-.5) {};

	\node (3) at (6,.5) {};
	\node (4) at (6,0) {};
	\node (5) at (6,-.5) {};

	\node[labeled] (v1) at (0, 1.5) {$v_1$};
	\node[labeled] (v2) at (4, -1.5) {$v_2$};

	\draw[very thick] (0) -- (1) -- (2) edge[out=-12,in=12,distance=10cm] (0)  (w1) -- (1) -- (w2) -- (2) -- (w1) -- (0) -- (w2);
	\draw[very thick] (3) -- (4) -- (5) edge[out=-12,in=12,distance=35mm] (3)  (w2) -- (4) -- (w3) -- (5) -- (w2) -- (3) -- (w3);
	\draw[dashed] (v1) edge[out=-12, in=90] (w2) (v1) edge[out=0, in=90] (w3);
	\draw[dashed] (v2) -- (w2) (v2) edge[out=0, in=-90] (w3);
	\draw (w1) edge[out=-90, in=180] (v2);

	\fill[color=black!10!white] (v1.center) to[bend right=120, distance=24mm] (w1.center) to (v1.center);
	\draw (v1.center) to[bend right=120, distance=24mm] (w1.center);

	\node[labeled] (xw1) at (w1) {$w_1$};
	\node[labeled] (xv1) at (v1) {$v_1$};
	\draw (v1) -- (w1);

	\node[draw=none, fill=none] (k8) at ($.5*(v1)+.5*(w1)-(.7,0)$) {$K_8$};
	\end{tikzpicture}
	\caption{Input for proving the strength of the pseudo-tree model. Expensive edges are bold. Removing the dashed edges allows to keep all expensive ones.}
	\label{fig:pseudo-tree-strength}
\end{figure}
 Consider the following input~$G$: start with two $K_5$'s that we each delete an arbitrary edge from and join them by identifying two degree-$3$ nodes with each other.
 We call this node~$w_2$.
 The resulting graph has exactly $3$ nodes~$W := \{w_1,w_2,w_3\}$ of degree not~$4$. We denote its edge-set by $E_5$.
 We add the nodes $v_1,v_2$ and connect both to all nodes of $W$.
 Let~$X$ denote a set of $6$ new nodes: we complete a $K_8$-subgraph on $X \cup \{w_1,v_1\}$ and denote its edges by $E_8$.
 Finally, we assign weight~$w$ to the edges: fix a (large) constant~$M \in \mathbb{N}$;
  all edges of $E_5$ have weight~$M$, all other edges have weight~$1$.
 See Fig.~\ref{fig:pseudo-tree-strength} for a schematic depiction of $G$.

 Consider $\CM 3$ on $G$. We show the fact for $\lpobj = 7$.
 We observe that all Kuratowski constraints are satisfied by the following $s$-variable assignment:
  $s_e = 0$ for all $e \in E_5 \cup \{v_2w_1\}$, $s_{v_1w_2} = s_{v_1w_3} = 17 / 18$, $s_{v_2w_2} = s_{v_2w_3} = 1$, and $s_e = 1 / 9$ otherwise.
 This gives an objective value of~$7$, independent of $M$.
 For this objective, the cycle constraint~\eqref{eq:ub} is satisfied if $c(3) \geq 28$ which is obtained by
  setting $c_\alpha = 1$ for $12$ triangles~$\alpha$ in $E_5$ (corresponding to an actual embedding of the two joined $K_5$-subgraphs) and $c_\alpha = 8 / 27$ for all $8 \choose 3$ triangles~$\alpha$ in $E_8$
  without violating constraints~\eqref{eq:at-most-2-cycles}.
 Note that $\deg(v_2)=1$ in the solution.

 Consider either variant  of the pseudo-tree model~\eqref{eq:pc-one-arc}--\eqref{eq:ubx},
  assume that the objective value would remain feasible.
 We observe that for any two edges $e,f \in \delta_G(v_2)$, the graph~$G[E_5] + e + f$ is not planar.
 For any $6$ nodes in $V(G[E_8])$ we obtain a $K_{3,3}$-subdivision. Summing over the respective Kuratowski constraints, it follows that $s(E_8) \geq |E_8| / 9$.
 Note that no edge in $E(G) \setminus (E_5 \cup E_8)$ is part of a triangle.
 Let $c(F)$ denote the sum of triangles that use edges of $F \subseteq E(G)$.
 By the extended cycle constraint~\eqref{eq:ubx} we have $c(3) \geq 28 + 2 t_{v_2}$.
 Since $c(3) \leq c(E_5) + c(E_8) \leq 2 / 3 \big(|E_5| + |E_8|-s(E_8) \big) \leq 2 / 3 \big(18 + 28 - 28 / 9 \big) = 772 / 27$,
 we obtain $t_{v_2} \leq 8 / 27$. Constraints~\eqref{eq:pc-prop-node} imply that $s\big(\delta_G(v_2)\big) \leq t_{v_2} + 1 = 35 / 27$.
 Hence, there exists a pair of edges~$e,f \in \delta_G(v_2)$ such that $s_e+s_f \leq 35 / 27 \cdot 2 / 3 = 70 / 81$ and $s(E_5) \geq 11 / 81$ follows.
 By choosing $M > (7 \cdot 81) / 11$ we obtain a contradiction.
\end{proof}

All following constraint classes deal with excluding combinations of cycles and paths that either induce non-planarity, or
 result in cycle-variables not assignable to any face in the planar subgraph (Lemma~\ref{lem:cycle-vars-faces}).
They are independent of but compatible with the pseudo-tree extension.

\subparagraph*{Cycle-Edge Constraints.}
 Considering integral solutions and constraints~\eqref{eq:at-most-2-cycles}, a cycle cannot be picked if any of its edges is deleted.
 W.r.t.\ fractional solutions we can additionally require
 \begin{align}
  s_e + c_\alpha &\leq 1 & \forall \alpha \in \mathcal C_{\leq D}, e \in E(\alpha) \label{eq:del-cycle}.
 \end{align}
 Although there are only $\bigO(D |\mathcal C_{\leq D}|)$ such constraints,
 preliminary benchmarks showed that adding all of them does not pay off.
 Instead, we straight-forwardly  separate them by iterating over the edges of each cycle
  that has a non-zero variable.

\begin{lemmarep}\atA\label{lem:cestr}
 The cycle-edge constraints~\eqref{eq:del-cycle} strengthen \CM 3.
\end{lemmarep}

\begin{proofsketch}
 Use a graph (Fig.~\ref{sfg:cycle-edge-strength}) that has $3$~edges each incident to only 1~triangle.
\end{proofsketch}

\begin{proof}
\begin{figure}\centering
	\begin{tikzpicture}[every node/.style=vertex]
	\path[use as bounding box] (-3,-2.7) rectangle (4,2.7);
	\coordinate (c_v1) at (330:1.5);
	\coordinate (c_v2) at (210:1.5);
	\coordinate (c_v3) at (90:1.5);

	\coordinate (c_w1) at (-20:.5);
	\coordinate (c_w2) at (200:.5);

	\fill[color=black!10!white] (c_v1) -- (c_w1) -- (c_w2) -- (c_v2) -- cycle;
	\fill[pattern=north west lines, pattern color=black!50!white] (c_v1) -- (c_w1) -- (c_w2) -- (c_v2) -- (c_v3) -- cycle;

	\node[labeled] (v1) at (c_v1) {$v_1$};
	\node[labeled] (v2) at (c_v2) {$v_2$};
	\node[labeled] (v3) at (c_v3) {$v_3$};

	\node (v4) at (155:4) {};
	\node (v5) at (25:4) {};
	\node (v6) at (-25:4) {};
	\node (v7) at (-155:4) {};

	\draw (v1) -- (v2) -- (v3) -- (v1);

	\draw (v4) edge[out=10,in=170] (v5) (v5) -- (v6) (v6) edge[out=190,in=-10] (v7) (v7) -- (v4);
	\draw (v4) edge[dashed, out=25, in=45, distance=5cm] (v6);
	\draw (v5) edge[out=-45, in=-25, distance=5cm] (v7);

	\draw (v4) -- (v3) (v4) edge[dashed] (v2) (v4) edge[dashed, out=-80, in=-125] (v1);
	\draw (v5) edge[dashed] (v1) (v5) -- (v3) (v5) edge[dashed, out=-100, in=-80] (v2);
	\draw (v6) -- (v1) (v6) -- (v2) (v6) -- (v3);
	\draw (v7) edge[dashed] (v1) (v7) -- (v2) (v7) -- (v3);

	\node[labeled] (w1) at (c_w1) {$w_1$};
	\node[labeled] (w2) at (c_w2) {$w_2$};

	\draw[very thick] (v1) -- (w1)  -- (w2) -- (v2);
	\draw (w1) -- (v3) -- (w2);

	\node[draw=none] (q) at (-90:.45) {$Q$};
	\node[draw=none] (t) at (90:.3) {$T$};
	\end{tikzpicture}
	\caption{Input for proving strength of cycle-edge constraints.
		The solid edges induce a planar subgraph and show that the same skewness is maintained when inserting $w_1,w_2$ into the $K_7$.
		Quadrangle~$Q$ is shaded and the set~$T$ of triangles is hatched.
		All edges incident with exactly one triangle are drawn in bold.}
	\label{fig:cycle-edge-strength}
\end{figure}
 Consider the $K_7$, and pick three nodes $v_1,v_2,v_3 \in V(K_7)$.
 We add the new nodes $w_1,w_2 \not \in V(K_7)$ and new edges $w_1w_2,v_1w_1,v_2w_2,v_3w_1,v_3w_2$ to obtain $G$, see Fig.~\ref{fig:cycle-edge-strength}.
 Note that the edges $I := w_1w_2,v_1w_1,v_2w_2$ are incident with exactly one triangle
 (each with a different one, let $T$ denote the set of the three triangles) and a shared quadrangle $Q=v_1w_1w_2v_2$.
 We observe that $\skewness(G) = \skewness(K_7) = 6$.

 Consider $\CM 3$ on $G$. We show the fact for $\lpobj = 5$.
 We choose $s_e = 1 / 8$ for all edges~$e \not \in E(Q)$ and set $s_{v_1w_1} = s_{v_2w_2} = 1 / 2$, $s_{v_1v_2} = s_{w_1w_2} = 5 / 8$.
 Without constraints~\eqref{eq:del-cycle},
  this allows us to set $c(3)=14$ (each edge not in~$E(Q)$ being incident with two triangles~$\alpha,\beta$
  whose cycle variables are $1$ and $3/4$, respectively),
  satisfying the cycle constraints.
 It is straight-forward to verify the existence of this cycle variable assignment (but manually tedious).
 Clearly, this bound is no better than the trivial Euler constraint on~$G$.

 However, when adding the cycle-edge constraints~\eqref{eq:del-cycle}, this solution becomes infeasible: for example $s_{v_1w_1} = 1/2$ but $c_{v_1w_1v_3} \geq 3 / 4$.
 We show that the objective is larger than $5$ by contradiction.
 To obtain an objective value of $5$, we require $c(3) \geq 14$.
 Let $E_7$ denote the edges in the $K_7$-subgraph of $G$, $\bar E_7 := E(G) \setminus E_7$, and $x := s(E_7)$.
 Recall that $s(E(G)) \geq 5$ by Euler and hence $s(\bar E_7) \geq 5-x$.
 Let $c(F)$ denote the sum of triangles where each triangle is weighted by its number of edges in $F \subseteq E(G)$ divided by $3$.
 Already by summing up constraints~\eqref{eq:at-most-2-cycles} over all edges of $E_7$
  we have $c(E_7) \leq 2 / 3 \big(|E_7|-x\big) = 14 - 2 / 3 x$ and hence $c(\bar E_7) \geq 2 / 3 x$.
 Note that $c(\bar E_7) = \sum_{\alpha \in T} c_\alpha$.
 For each $\alpha \in T$ with $c_\alpha > 0$,
  we require a certain number of edges in $\bar E_7$ not to be deleted.
 More precisely, we may assume the largest two cycle variables of $T$ to be incident with $4$ different edges of $\bar E_7$
  while the smallest one adds just a single edge of $\bar E_7$ that is not used in the other two cycles.
 Assuming the new constraints to be satisfied on $I$, we obtain $s(\bar E_7) \leq |\bar E_7| - c(\bar E_7)(2 / 3 \cdot 2 + 1 / 3)$.
 Since $K_7$ contains a Kuratowski-subdivision, it follows that $x > 0$ and we have the contradiction $5 - x \leq s(\bar E_7) \leq 5 - 10 / 9 x$.
\end{proof}

\insertProofSketchFigure

\subparagraph*{Two-Cycles-Path Constraints.}
Given two cycles~$\alpha,\beta$, we denote their set of inner nodes by $\nu(\alpha,\beta) := \{ v \in V(\alpha) \cap V(\beta) \mid \delta_\alpha(v) = \delta_\beta(v) \}$.
Let $\psi(\alpha,\beta)$ denote the set of non-empty paths that connect $\nu(\alpha,\beta)$ to $ V(\alpha \sqcup \beta)$ without using any edge in $E(\alpha \sqcup \beta)$.
\begin{lemma}
 The two-cycles-path constraints, given below, are feasible.
 \begin{align}
  s\big(E(p)\big) &\geq c_{\alpha} + c_{\beta} - 1 & \forall \alpha,\beta \in \mathcal C_{\leq D}; p \in \psi(\alpha, \beta) \label{eq:two-cycles-path}
 \end{align}
 \begin{proof}
  Assume an embedding~$\pi$ of~$\alpha \sqcup \beta$ where
   each of $\alpha,\beta$ corresponds to a face in $\pi$.
  By inserting $p$ into $\pi$, we either split face $\alpha$ or face $\beta$.
  Hence, even in a supergraph of $\alpha \sqcup \beta \sqcup p$ two such faces cannot exist.
  Otherwise, if no such $\pi$ exists, we have $1 \geq c_{\alpha} + c_{\beta}$.
 \end{proof}
\end{lemma}

\begin{lemmarep}\atA\label{lem:twocpstr}
 The two-cycles-path constraints~\eqref{eq:two-cycles-path} strengthen \CM 4.
\end{lemmarep}

\begin{proofsketch}
 We use the graph of Fig.~\ref{sfg:two-cycles-path-strength} as input.
\end{proofsketch}

\begin{proof}
 \begin{figure}\centering
	\begin{subfigure}[t]{.44\textwidth}\centering
		\begin{tikzpicture}[every node/.style=vertex, scale=.6]
		\node[labeled] (w) at (0,0) {$w$};
		\node (0) at (150:3) {};
		\node (1) at (90:4) {};
		\node[labeled] (v1) at (90:3) {$v_1$};
		\node (2) at (90:2) {};
		\node (3) at (30:3) {};
		\node (4) at (-30:4) {};
		\node[labeled] (v2) at (-30:3) {$v_2$};
		\node (5) at (-30:2) {};
		\node (6) at (-90:3) {};
		\node (7) at (-150:4) {};
		\node[labeled] (v3) at (-150:3) {$v_3$};
		\node (8) at (-150:2) {};

		\draw[bend left] (0) edge (1) (1) edge (3) (3) edge (4) (4) edge (6) (6) edge (7) (7) edge (0);
		\draw (2) edge (3) (3) edge (5) (5) edge (6) (6) edge (8) (8) edge (0);
		\draw (1) -- (v1) -- (2) (4) -- (v2) -- (5) (7) -- (v3) -- (8);

		\fill[color=black!10!white] (v1.center) to[bend right=90, distance=26mm] (w.center) to[bend left] (v1.center);
		\draw[bend left] (v1.center) to[bend right=90, distance=26mm] (w.center) to (v1.center) (w.center) edge[dashed] (v2.center) (w.center) edge[dashed] (v3.center);

		\node[labeled] (xw) at (0,0) {$w$};
		\node[labeled] (xv1) at (90:3) {$v_1$};
		\node[labeled] (xv2) at (-30:3) {$v_2$};
		\node[labeled] (xv3) at (-150:3) {$v_3$};
		\draw (0) edge[bend right=60] (2);

		\node[draw=none] (k8) at (125:1.9) {$K_8$};
		\end{tikzpicture}
		\caption{The input~$G$ with a schematic depiction of the $K_8$-subgraph. $G[V(G) \setminus X]$ without the dashed edges is planar.}
		\label{sfg:tcps-input}
	\end{subfigure}\hfill
	\begin{subfigure}[t]{.52\textwidth}\centering
		\begin{tikzpicture}[every node/.style=vertex, scale=.6]
		\node (0) at (150:3) {};
		\node (1) at (90:4) {};
		\node[labeled] (v1) at (90:3) {$v_1$};
		\node (2) at (90:2) {};
		\node (3) at (30:3) {};
		\node (4) at (-30:4) {};
		\node[labeled] (v2) at (-30:3) {$v_2$};
		\node (5) at (-30:2) {};
		\node (6) at (-90:3) {};
		\node (7) at (-150:4) {};
		\node[labeled] (v3) at (-150:3) {$v_3$};
		\node (8) at (-150:2) {};

		\fill[color=black!10!white] (0.center) to[bend left] (1.center) to[bend left] (3.center) to (2.center) to cycle;
		\fill[pattern=north west lines, pattern color=black!65!white] (0.center) to[bend left] (1.center) to (2.center) to cycle;

		\draw[bend left] (0.center) to (1.center) to (3.center);

		\node[labeled] (x) at (0) {$x$};
		\node[fill=white] (l1) at (1) {};
		\node[labeled] (lv1) at (v1) {$v_1$};
		\node[fill=white] (l2) at (2) {};
		\node[labeled] (y) at (3) {$y$};

		\draw[bend left] (y) edge[very thick] (4) (4) edge (6) (6) edge (7) (7) edge (x);
		\draw (x) -- (2) edge (y) (y) edge (5) (5) edge[very thick] (6) (6) edge[very  thick] (8) (8) edge[very thick] (x);
		\draw (1) -- (v1) -- (2) (4) edge[very thick] (v2) (v2) edge[very thick] (5) (7) -- (v3) -- (8);
		\end{tikzpicture}
		\caption{Example of a two-cycles-path constraint: if no bold edge is deleted,
			the shaded and the (overlapping) hatched quadrangle cannot both occur as part of a face (here, only the hatched one does).}
		\label{sfg:tcps-constraint}
	\end{subfigure}
	\caption{Graphs for proving strength of two-cycles-path constraints.}
	\label{fig:two-cycles-path-strength}
\end{figure}
 We construct our input~$G$ in the following manner, see Fig.~\ref{sfg:tcps-input}:
 consider the $K_3$ and replace each of the three edges by a new $K_{2,3}$ such that its end nodes become
 two nodes of the cardinality-$3$ node partition of $K_{2,3}$.
 Let $v_1,v_2,v_3$ denote the nodes of the cardinality-$3$ partitions that are not identified with any of $V(K_3)$.
 We add a new node $w$ and the edges $wv_2,wv_3$.
 Finally, we add six new nodes~$X$ and a $K_8$-subgraph on $X \cup \{w,v_1\}$.
 Let $E_8$ denote the edge set of this $K_8$-subgraph.
 Note that $w$ is not incident with any quadrangle outside of $E_8$ and all triangles of $G$ are contained in $E_8$.

 Consider $\CM 4$ on $G$. We show the fact for $\lpobj = 6$.
 First, note that the graph~$G - E_8$ becomes planar when removing any edge incident with $w$.
 In fact, all Kuratowski constraints are satisfied by setting $s_e = 1 / 9$ for all $e \in E_8$ and $s_{wv_2} = s_{wv_3} = 1$.
 Let $F := E(G) \setminus \big(E_8 \cup \delta_G(w)\big)$.
 To obtain an LP-feasible $c$-variable assignment, we additionally set $s_e = 4 / 81$ for all edges~$e \in F$,
  set $c_\alpha = 8 / 27$ for all triangles~$\alpha$ in $E_8$, and
  set $c_\alpha = 77 / 81$ for all $9$~quadrangles~$\alpha$ in $F$.
 It is easy to see that constraints~\eqref{eq:at-most-2-cycles} and \eqref{eq:ub} are satisfied.
 Hence, an objective value of~$6$ is feasible unless the new constraints are added.

 We now show that this is no longer the case when adding the new two-cycles-path constraints~\eqref{eq:two-cycles-path}.
 Assume otherwise, i.e., $s(E(G)) \leq 6$.
 We denote the sum over all variables of quadrangles in $F$ that are not incident with any $v_i$ ($i \in \{1,2,3\}$) by $c_\mathrm{out}$.
 Similarly, we denote the sum over all variables of quadrangles in $F$ that are incident with some $v_i$ ($i \in \{1,2,3\}$), by $c_\mathrm{in}$
 and the sum over all remaining quadrangle variables (each fully contained in $E_8$) by $c_8(4)$.
 The Kuratowski constraints on $E_8$ imply $s(E_8) \geq |E_8|/9$ and
  it follows from constraints~\eqref{eq:at-most-2-cycles} that $3 / 2 \cdot c(3) + 4 / 2 \cdot c_8(4) \leq |E_8|-s(E_8)$.
 Hence, $c(3) + c_8(4) \leq 448 / 27$.
 The cycle constraint~\eqref{eq:ub} implies $41 \leq 2c(3) + c(4)$.

 Let us apply the new constraints:
  for any~$i \in \{1,2,3\}$, pick node $v_i$ and one of its (two) incident quadrangles~$Q_i$ that we denote by~$\alpha_\mathrm{in} \in Q_i$.
 There is a unique quadrangle~$\alpha_\mathrm{out}$ not incident with $v_i$ but with two $v_i$'s $F$-neighbors.
 Let $x \in V(\alpha_\mathrm{in}) \cap V(\alpha_\mathrm{out})$ denote the unique node with $\delta_{\alpha_\mathrm{in}}(x) = \delta_{\alpha_\mathrm{out}}(x)$,
  i.e.,  the inner node of $\alpha_\mathrm{in}$ with $\alpha_\mathrm{out}$.
 We denote the single node contained in $V(\alpha_\mathrm{out}) \setminus V(\alpha_\mathrm{in})$ by~$y$.
 We observe that any $x$-$y$-path~$p$ in $G - E_8 - E(\alpha_\mathrm{out})$ forms a two-cycle-path constraint with $\alpha_\mathrm{out}$ and $\alpha_\mathrm{in}$, see Fig.~\ref{sfg:tcps-constraint}.
 Summing over all such constraints (there are exactly $16$~paths for two fixed cycles), we obtain $2s(F) \geq 2c_\mathrm{out}+c_\mathrm{in}-6$.
 By constraints~\eqref{eq:at-most-2-cycles}, we have $c_\mathrm{out}+c_\mathrm{in} \leq 9 - s(F) / 2$, an upper bound on $s(F)$.
 Combining both we obtain $30 \geq 6c_\mathrm{out} + 5c_\mathrm{in}$ and hence $c_\mathrm{out} + c_\mathrm{in} \leq 6$.
 We finally obtain $41 \leq 2c(3) + c(4) \leq 2\big( 448/27-c_8(4)\big) + c_8(4) + c_\mathrm{out} + c_\mathrm{in} \leq 2 \cdot 448 / 27 + 6 < 40$, a contradiction.
\end{proof}

To identify violated two-cycles-path constraints, we consider each edge $e$.
 We collect the set~$C(e) = \{\alpha \in \mathcal{C}_{\leq D} \mid e \in E(\alpha) \wedge c_\alpha > 0 \}$, and check,
 for each pair $\alpha,\beta\in C(e)$, whether
 its sum of LP-values is $>1$.
If so, we compute the set of inner nodes $\nu:=\nu(\alpha,\beta)$ and cache the result for future lookup.
If $\nu \neq \varnothing$, we iteratively compute shortest paths following either of two patterns:
the \emph{combined} approach searches for shortest paths from $\nu$ to $V(\alpha \sqcup \beta) \setminus \nu$,
whereas the \emph{separate} one searches for paths from $v$ to $V(\alpha \sqcup \beta) \setminus \{v\}$, separately for each $v \in \nu$.
Note that the latter variant will always identify a violated constraint, if one exists,
 whereas the former ignores paths connecting two inner nodes.
After identifying a new path~$p$, an edge in $E(p)$ with maximal LP-value is discarded and the search at $v$ starts anew.

We point out that there is a natural generalization of this constraint class by using $k$ instead of only $2$ cycles.
If the $k$ cycles fully enclose a common node~$v$ (like any $2$ cycles enclose their inner nodes), any other path from $v$
 to the same block is forbidden.

\subparagraph*{Cycle-Two-Paths Constraints.}
We say that two paths~$p_1,p_2$ are \emph{conflicting w.r.t.\ a cycle}~$\alpha$ if and only if
 they each start and end on nodes of $V(\alpha)$ but are otherwise disjoint from $\alpha$ and from one another, and
 $p_2$ connects the components of $\alpha[V(\alpha) \setminus V(p_1)]$.
\begin{lemma}
 The cycle-two-paths constraints, given below, are feasible.
 \begin{align}
  s\big(E(p_1 \sqcup p_2)\big) &\geq c_\alpha & \forall \alpha \in \mathcal C_{\leq D}, \forall\text{ conflicting paths }p_1, p_2\text{ w.r.t.\ } \alpha \label{eq:cycle-two-paths}
 \end{align}
 \begin{proof}
  Given an embedding~$\pi$ of $\alpha$, we cannot insert both paths~$p_1,p_2$ into the same face of~$\pi$.
  Hence, we must split both faces in $\pi$. Consequently,
  no embedding of any supergraph of $\alpha \sqcup p_1 \sqcup p_2$ exists, where there is a face incident with all of $\alpha$.
 \end{proof}
\end{lemma}

While this constraint class may be stronger than the two-cycles-path constraints,
 we did not implement it: its separation is complex
 as we ask for two paths depending on each other.

\subparagraph*{Kuratowski-Cycle Constraints.}
Starting with a Kuratowski constraint, we can replace parts of its edges by cycles that contain them.
\begin{lemma}
 The Kuratowski-cycle constraints, given below, are feasible.
 \begin{align}
  s(\{e \in E(K) \mid \forall \alpha \in C : e \not\in E(\alpha)\}) &\geq \sum_{\alpha \in \mathcal C}c_\alpha + 1 - |C| & \forall K \in \mathcal K, C \subseteq \mathcal C_{\leq D} \label{eq:cycle-kurat}
 \end{align}
 \begin{proof}
  If $C=\varnothing$, we simply obtain a Kuratowski constraint.
  Assuming integrality and $C \neq \varnothing$, the right-hand side is $1$ if all cycles in $C$ are picked and $\leq 0$ otherwise.
  In the former case, the edges of $C$, together with the remaining edges of $K$ that are not contained in $C$ contain a Kuratowski subdivision, and we need to remove an edge.
 \end{proof}
\end{lemma}

\begin{lemmarep}\atA
 The Kuratowski-cycle constraints~\eqref{eq:cycle-kurat} strengthen \CM 4.
\end{lemmarep}

\begin{proofsketch}
 We use the circulant on $16$~nodes with jumps~$1$, $2$, and~$8$ as input.
\end{proofsketch}

\begin{proof}
\begin{figure}\centering
	\begin{subfigure}[t]{.3\textwidth}\centering
		\begin{tikzpicture}[every node/.style=vertex,scale=.7]
		\path[use as bounding box] (-3,-3) rectangle (3,3);

		\foreach \i [evaluate=\i as \x using \i*22.5] in {0,...,15}{
			\node (\i) at (\x:2.5) {};
		}

		\draw (0) -- (1) -- (2) -- (3) -- (4) -- (5) -- (6) -- (7) -- (8) -- (9) -- (10) -- (11) -- (12) -- (13) -- (14) -- (15) -- (0);
		\draw[bend right=70] (0) edge (2) (2) edge (4) (4) edge (6) (6) edge (8) (8) edge (10) (10) edge (12) (12) edge (14) (14) edge (0);
		\draw[bend left] (1) edge (3) (3) edge (5) (5) edge (7) (7) edge (9) (9) edge (11) (11) edge (13) (13) edge (15) (15) edge (1);

		\newcommand{\dash}[2]{
		 \pgfmathsetmacro\result{(8-#1)*0.2+4}
		 \draw[dash pattern=on \result pt off 2pt] (#1) edge (#2);
		}
		\dash 0 8
		\dash 1 9
		\dash 2 {10}
		\dash 3 {11}
		\dash 4 {12}
		\dash 5 {13}
		\dash 6 {14}
		\dash 7 {15}
		\end{tikzpicture}
		\caption{Input: circulant graph with jumps~$1$ (solid straight), $2$ (solid bent), and $8$ (dashed):
			adding any single edge of $E_8$ to $G[E_1 \cup E_2]$ maintains in a planar graph.
		}
		\label{sfg:kcs-input}
	\end{subfigure}\hfill%
	\begin{subfigure}[t]{.3\textwidth}\centering
		\begin{tikzpicture}[every node/.style=vertex,scale=.7]
		\path[use as bounding box] (-3,-3) rectangle (3,3);

		\foreach \i [evaluate=\i as \x using \i*22.5] in {0,...,15}{
			\node (\i) at (\x:2.5) {};
		}

		\fill[color=black!10!white] (5.center) to[bend right=70] (3.center) to (11.center) to[bend left=70] (13.center) to (5.center);
		\draw[bend left=70] (3.center) to (5.center) (11.center) to (13.center);

		\newcommand{\tri}[3]{
			\fill[pattern=north west lines, pattern color=black!50!white] (#1.center) to[bend right=80, distance=1cm] (#3.center) to (#2.center) to (#1.center);
			\draw[bend right=80, distance=1cm] (#1.center) to (#3.center);
		}

		\tri012
		\tri234
		\tri456
		\tri678
		\tri89{10}
		\tri{10}{11}{12}
		\tri{12}{13}{14}
		\tri{14}{15}0

		\foreach \i [evaluate=\i as \x using \i*22.5] in {0,...,15}{
			\node[fill=white] (xx\i) at (\x:2.5) {};
		}

		\node[fill=black!30!white] (v2) at (2) {};
		\node[fill=black!10!white] (v3) at (3) {};
		\node (v4) at (4) {};
		\node[fill=black!30!white] (v5) at (5) {};
		\node[fill=black!10!white] (v6) at (6) {};
		\node[fill=black!30!white] (v11) at (11) {};
		\node[fill=black!10!white] (v13) at (13) {};

		\draw (0) -- (1) -- (2) -- (3) (3) -- (4)  (4) -- (5) (5) -- (6) -- (7) -- (8) -- (9) -- (10) -- (11) -- (12) -- (13) -- (14) -- (15) -- (0);
		\draw (3) -- (11) (5) -- (13);
		\end{tikzpicture}
		\caption{This subgraph contains a $K_{3,3}$-subdivision and consists of
			eight triangles (hatched) and a $2$-$8$-quadrangle (shaded).}
		\label{sfg:kcs-subdiv-b}
	\end{subfigure}\hfill%
	\begin{subfigure}[t]{.3\textwidth}\centering
		\begin{tikzpicture}[every node/.style=vertex,scale=.7]
		\path[use as bounding box] (-3,-3) rectangle (3,3);

		\foreach \i [evaluate=\i as \x using \i*22.5] in {0,...,15}{
			\node (\i) at (\x:2.5) {};
		}

		\fill[color=black!10!white] (5.center) to[bend right=70] (3.center) to (11.center) to[bend left=70] (13.center) to (5.center);
		\draw[bend left=70] (3.center) to (5.center) (11.center) to (13.center);

		\node[fill=black!10!white] (v3) at (3) {};
		\node[fill=black!30!white] (v5) at (5) {};
		\node[fill=black!30!white] (v11) at (11) {};
		\node[fill=black!10!white] (v13) at (13) {};

		\draw (8) -- (0);

		\node[fill=black!30!white] (v0) at (0) {};
		\node[fill=black!10!white] (v8) at (8) {};

		\draw (0) -- (1) -- (2) -- (3) (3) (5) (5) -- (6) -- (7) -- (8) -- (9) -- (10) -- (11) (13) -- (14) -- (15) -- (0);
		\draw (3) -- (11) (5) -- (13);
		\end{tikzpicture}
		\caption{$K_{3,3}$-subdivision that consists of $12$ edges of $E_1$, one edge of $E_8$ and a disjoint $2$-$8$-quadrangle (shaded).}
		\label{sfg:kcs-subdiv-c}
	\end{subfigure}
	\caption{Graphs for showing the strength of Kuratowski-cycle constraints.}
	\label{fig:gen-kurat-cycle-strength}
\end{figure}
 Let $G$ denote the circulant graph on $16$~nodes with jumps~$1,2$, and~$8$ (see Fig.~\ref{sfg:kcs-input}).

 Consider $\CM 4$ on $G$.  We show the fact for $\lpobj = 14 / 3$.
 Assume the nodes of $G$ to be labeled as $v_0,v_1,\ldots,v_{15}$, corresponding to jumps~$1$.
 Let $E_i \subseteq E(G)$ denote the edges corresponding to jumps~$i$.
 We observe that every Kuratowski subdivision needs at least two edges of $E_8$ as $(V(G),E_1 \cup E_2\cup\{e\})$ is planar for all $e \in E_8$ (see Fig.~\ref{sfg:kcs-input}).
 It follows that every Kuratowski constraint is satisfied by $s_e = 1 / 2$ for all $e \in E_8$.
 Without the new constraints, an LP-value of $14 / 3$ is obtainable by
  $s_e = 1 / 2$ for all $e \in E_8 \cup \{s_{v_1v_3}\}$, $s_{v_5v_7} = 1 / 6$, and $s_e = 0$ otherwise; and
  by picking all $16$ triangles and four disjoint quadrangles~$R$,
  consisting of two edges of $E_2 \setminus \{v_1v_3, v_5v_7\}$ and $E_8$ each,
  with cycle variable value~$1$.

 The solution is cut-off by the new Kuratowski-cycle constraints~\eqref{eq:cycle-kurat}%
:
 intuitively, we may pick a quadrangle of $R$ instead of two edges of $E_8$ to obtain a violated Kuratowski-cycle constraint
 although the sum of any two such edges being~$1$ prevents any traditional Kuratowski constraints to be violated.

 We now show that $s(E(G)) > 14 / 3$ when adding a certain set of Kuratowski-cycle constraints.
 An $i$-$j$-quadrangle is a quadrangle that uses edges of $E_i$ and $E_j$.
 Let $q_{ij}$ denote the sum over variables corresponding to $i$-$j$-quadrangles.
 Observe that $c(4) = q_{12} + q_{18} + q_{28}$, since there are no other quadrangles in~$G$.
 The cycle constraint~\eqref{eq:ub} implies that $3s(E(G)) + 2c(3) + q_{12} + q_{18} + q_{28} \geq 50$.
 By summing over all constraints~\eqref{eq:at-most-2-cycles} that correspond to edges of $E_1$, we obtain
  $c(3) + q_{12} + q_{18} + s(E_1) \leq 16$. Similarly, using $E_8$ instead of $E_1$,
  we have $q_{18} + q_{28} + s(E_8) \leq 8$.

 We describe three types of Kuratowski-cycle constraints~\eqref{eq:cycle-kurat} in~$G$.
 Let $W$ denote the nodes of $G$ that have an odd index. We observe that $G[W]$ is a M\"obius ladder:
  it consists of exactly four $2$-$8$-quadrangles and is non-planar. The same construction works for even node indices.
 We obtain $q \leq 6$ by summing over both constraints.
 Consider a single $2$-$8$-quadrangle~$\alpha$: the indices of all nodes in~$\alpha$ have identical parity~$p$.
 The graph that consists of $\alpha$ and the $8$~triangles that use edges of $E_2$ incident with nodes of parity not~$p$, is non-planar, see Fig.~\ref{sfg:kcs-subdiv-b}.
 Summing over all corresponding Kuratowski-cycle constraints, we obtain $64 \geq 4c(3) + q_{28}$.
 Once again, consider a single $2$-$8$-quadrangle~$\alpha$:
  there are exactly two nodes~$u,v \not \in V(\alpha)$ each incident with two nodes of $V(\alpha)$.
 Pick any $f \in E_8 \cap E(\alpha)$ such that $f \cap \{u,v\} = \varnothing$.
 The graph $\alpha \sqcup G[f \cup \{e \in E_1 \mid e \cap \{u,v\} = \varnothing \}]$ is non-planar, see Fig.~\ref{sfg:kcs-subdiv-c}.
 It consists of a $2$-$8$-quadrangle, another edge of $E_8$, and $12$~edges of $E_1$.
 Summing over all corresponding Kuratowski-cycle constraints, we have $6s(E_1) + s(E_8) \geq q_{28}$.

 By the above, we obtain a system of linear inequalities on six non-negative variables:
 \begin{align*}
  s(E_1) + s(E_2) + s(E_8) &\leq 14 / 3 \\
  3s(E_1) + 3s(E_2) + 3s(E_8) + 2c(3) + q_{12} + q_{18} + q_{28} &\geq 50 \\
  c(3) + q_{12} + q_{18} + s(E_1) &\leq 16 \\
  q_{18} + q_{28} + s(E_8) &\leq 8 \\
  q &\leq 6 \\
  4c(3) + q_{28} &\leq 64 \\
  6s(E_1) + s(E_8) &\geq q_{28}
 \end{align*}
 It is straight-forward (although manually tedious) to check that this system has no solution.
\end{proof}

For separation, we identify a Kuratowski subdivision $K$ as for~\eqref{eq:kurat}.
 We collect the set~$S$ of cycles with LP-value $>0$ incident with $K$.
For each cycle in~$S$, we compute its \emph{gain}, i.e., the increase in violation (or decrease in slack) when adding that cycle to $C$.
While there are cycles with positive gain, we continue adding a cycle of $S$ with maximal gain to $C$.

\subparagraph*{Cycle-Clique Constraints.}
Two cyclic orders~$\pi, \bar\pi$ on a set $X$ are \emph{conflicting} if and only if $\pi \neq \bar\pi$ and $\pi \neq \mathrm{reverse}(\bar\pi)$.
The restriction of~$\pi$ to $Y \subseteq X$ is denoted by $\pi^Y$.
A cycle$~\alpha$ induces a (up to reversal) unique cyclic order on its nodes~$V(\alpha)$.
Given two cycles~$\alpha,\beta$, let~$\pi_\alpha, \pi_\beta$ be corresponding cyclic orders, and let $W := V(\alpha) \cap V(\beta)$ be the common nodes.
We say that $\alpha$ and $\beta$ are \emph{conflicting} if and only if $\pi^W_\alpha$ and $\pi^W_\beta$ are conflicting.
\begin{lemma}
 The cycle-clique constraints, given below, are feasible.
 \begin{align}
  \sum_{\alpha \in C} c_\alpha &\leq 1 &\forall C \subseteq \mathcal C_{\leq_D} \text{ s.t.\ all cycles in $C$ are pairwise conflicting} \label{eq:cycle-clique}
 \end{align}
 \begin{proof}
  Consider any pair of conflicting cycles $\alpha,\beta\in C$ with $\pi_\alpha$, $\pi_\beta$, and $W$ defined as above.
  Since cyclic orders on three elements are unique up to reversal, we have $|W| \geq 4$.
  By transitivity there exists a set of exactly four common nodes $X \subseteq W$,
   such that $\pi^X_\alpha$ and $\pi^X_\beta$ are conflicting.
  The graph on $X$ where we add an edge~$vw$ if and only if $v$ is adjacent to $w$ in $\pi^X_\alpha$ or $\pi^X_\beta$ is the $K_4$.
  Since the $K_4$ is not outerplanar,
   there can neither be a face in $K_4$ traversing all of $X$ nor such a face in $\alpha \sqcup \beta$.
 \end{proof}
\end{lemma}

We create the conflict graph~$H_C$ that contains a node for every cycle with LP-value $>0$,
 cache the conflict information for each pair of cycles,
 and add constraints for maximal cliques in $H_C$.
In a less sophisticated variant, we only add constraints for cliques of size two.

\section{Experiments}\label{sec:experiments}
 All algorithms are implemented in \cpp, compiled with GCC 6.3.0, and use the OGDF (snapshot 2017-07-23)~\cite{OGDF}.
 We use SCIP 4.0.1 for solving ILPs with CPLEX 12.7.1 as the underlying LP solver~\cite{SCIP400}.
 Each MPS-computation uses a single physical core of a Xeon Gold 6134 CPU (3.2 GHz) with a memory speed of 2666 MHz.
 We employ a time limit of 20 minutes and a memory limit of 8 GB per computation.
 Our instances and results, giving runtime and skewness (if solved), are available for download at \texttt{http://tcs.uos.de/research/mps}.
 \todo{webseite anpassen: SEA paper zitieren + neues paper (data download)}

\subparagraph*{Instances and Algorithms.}
Analogously to the study~\cite{ChimaniHedtkeWiedera2018}, we consider
three established real-world benchmark sets:
 \emph{Rome}~\cite{DiBattistaGargLiottaTamassiaTassinariVargiu1997}, 
 \emph{North}~\cite{North1995},
 and (a subset of) \emph{SteinLib}~\cite{KMV00}.
We know from \cite{ChimaniHedtkeWiedera2016, ChimaniHedtkeWiedera2018} that random regular
graphs (which are \emph{expander graphs} with high probability) are especially hard to solve exactly.
We use the same such instances as~\cite{ChimaniHedtkeWiedera2018},
but only consider graphs with $\leq100$ nodes, as no known exact algorithm solves larger instances.
 There are 20 graphs for each parameterization~$(|V(G)|,\Delta) \in \{10,20,30,50,100\} \times \{4,6,10,20,40\}$,
 where $\Delta < |V(G)|$ is the node-degree.
For tuning of \void (e.g., heap size in separation) we rely on the values identified in~\cite{hedtkePhd}.
\newcommand{\tabpadding}{.5em}
We use the notation below to encode algorithmic choices.\vspace*{\tabpadding}\\
\begin{tabular}{|rl|}
 \hline
 \void & Do not use any extensions but the basic Kuratowski algorithm~\cite{Mutzel1994}.\\
 e & Separate generalized Euler constraints~\eqref{eq:gen-euler}.\\
 c\{$r$\} & Add cycle constraints~\eqref{eq:at-most-2-cycles},~\eqref{eq:ub}, and variables with the minimal value for $D$ such that\\
           & there are variables for at least $100r$ cycles.\\
 t\{0|1\} & Use the pseudo-tree extension~\eqref{eq:pc-one-arc}--\eqref{eq:ubx} with (=t1) or without (=t0) propagation.\\
 i & Enforce integrality of variables for cycles and pseudo-trees.\\
  s & Separate cycle-edge constraints~\eqref{eq:del-cycle}.\\
  w\{0|1\} & Separate two-cycles-path constraints~\eqref{eq:two-cycles-path} using \emph{combined} (=w0) or \emph{separate} (=w1)\\
           & approach. Also enables separation on cycle-clique constraints~\eqref{eq:cycle-clique} for $2$-cliques.\\
 k & Separate Kuratowski-cycle constraints~\eqref{eq:cycle-kurat}.\\
 q & Separate cycle-clique constraints~\eqref{eq:cycle-clique}.\\
 \hline
\end{tabular}\vspace*{\tabpadding}\\
Note that instead of providing~$D$ explicitly,
we specify a minimum number~$r'$ of cycle variables to be generated.
We increment~$D$ while there are less than $r'$~cycle variables.

\begin{table}[p]
\centering
\caption{Overview of performance for algorithmic variants: success rate and avg.\ runtime.}
\scalebox{\tablescale}{\begin{tabular}{|l|r|r|r|r|r|r|r|r|}
 \hline
 \multicolumn{1}{|c|}{variant} & \multicolumn{2}{c|}{Rome} & \multicolumn{2}{c|}{North} & \multicolumn{2}{c|}{Expanders} & \multicolumn{2}{c|}{SteinLib} \\
       & \ce{succ.\ [\%]} & \ce{time [s]} & \ce{succ.\ [\%]} & \ce{time [s]} & \ce{succ.\ [\%]} & \ce{time [s]} & \ce{succ.\ [\%]} & \ce{time [s]} \\
 \hline
 \void         &     85.71  &    198.42  &     73.76  &     325.31  &     34.74  &     800.38  &      9.52  &  1\,085.94  \\
 \hline
 e             &     85.56  &    199.41  &     77.78  &     273.29  &     35.00  &     803.91  &      9.52  &  1\,085.93  \\
 c5            &     98.91  &     21.60  &     84.40  &     201.42  &     53.95  &     567.52  &     31.43  &     859.40  \\
 c10           & \tb{99.14} & \tb{18.10} & \tb{84.63} & \tb{195.35} &     54.47  &     562.81  & \tb{32.38} & \tb{853.57} \\
 c20           & \tb{99.14} &     19.58  &     83.92  &     197.86  & \tb{55.00} &     573.82  &     31.43  &     861.47  \\
 \hline
 c10 i         &     99.89  &      5.52  &     88.89  &     156.99  &     56.58  &     538.81  &     31.43  &     841.88  \\
 c10 s         &     99.79  &      6.66  &     88.42  &     165.54  & \tb{58.68} & \tb{515.13} &     35.24  &     821.00  \\
 c10 t0        & \tb{99.95} & \tb{ 3.36} &     92.43  &     112.82  &     57.37  &     535.46  & \tb{37.14} &     789.26  \\
 c10 t1        &     99.92  &      3.74  & \tb{93.14} & \tb{111.94} &     56.32  &     539.04  & \tb{37.14} & \tb{785.89} \\
 c10 w0        &     99.79  &      7.07  &     87.23  &     165.36  &     55.53  &     549.94  &     31.43  &     837.52  \\
 c10 w1        &     99.82  &      6.63  &     86.52  &     179.48  &     55.00  &     553.38  &     31.43  &     833.81  \\
 c10 k         &     99.77  &      7.26  &     86.52  &     178.34  &     55.26  &     552.83  &     33.33  &     828.81  \\
 c10 q         &     99.73  &      7.51  &     85.82  &     185.84  &     55.53  &     550.55  &     31.43  &     841.77  \\
 \hline
 c10 t0 i      &     99.95  &      3.22  &     93.14  &     109.61  &     57.37  &     529.93  &     38.10  &     782.72  \\
 c10 t0 s      & \tb{99.98} &      3.08  & \tb{93.62} & \tb{ 95.46} & \tb{58.95} & \tb{509.01} & \tb{39.05} & \tb{760.70} \\
 c10 t0 w0     & \tb{99.98} & \tb{ 2.75} &     92.43  &     112.77  &     57.37  &     537.61  &     36.19  &     808.58  \\
 c10 t0 w1     & \tb{99.98} &      2.92  &     92.20  &     109.37  &     57.11  &     537.76  &     37.14  &     780.83  \\
 c10 t0 k      &     99.92  &      3.57  &     92.67  &     104.55  &     56.84  &     535.97  &     38.10  &     785.46  \\
 c10 t0 q      &     99.95  &      3.58  &     92.67  &     109.71  &     57.37  &     538.19  &     37.14  &     789.05  \\
 c10 t1 i      &     99.96  &      3.32  &     92.91  &     106.39  &     57.11  &     533.51  &     37.14  &     788.52  \\
 c10 t1 s      & \tb{99.98} & \tb{ 2.75} &     92.43  &     112.77  &     58.68  &     537.61  &     37.14  &     808.58  \\
 c10 t1 w0     & \tb{99.98} &      3.04  &     92.20  &     114.28  &     56.84  &     537.97  &     38.10  &     786.93  \\
 c10 t1 w1     & \tb{99.98} &      3.19  &     91.96  &     113.03  &     56.84  &     540.39  &     37.14  &     783.09  \\
 c10 t1 k      &     99.92  &      3.65  &     92.20  &     112.61  &     56.84  &     538.91  &     37.14  &     784.30  \\
 c10 t1 q      &     99.92  &      3.86  &     93.14  &     113.89  &     56.05  &     540.23  &     37.14  &     788.07  \\
 \hline
 c10 t0 i s    &     99.94  &      3.27  &     92.91  &     103.17  & \tb{58.95} &     506.63  & \tb{40.00} &     761.47  \\
 c10 t0 s w0   & \tb{99.98} &      2.43  & \tb{93.85} & \tb{ 91.66} &     58.68  &     508.28  &     39.05  &     763.08  \\
 c10 t0 s w1   & \tb{99.98} & \tb{ 2.29} &     92.91  &     101.17  &     58.68  &     507.99  &     39.05  &     756.31  \\
 c10 t0 s k    &     99.96  &      3.03  &     93.38  &      98.06  &     58.68  &     504.54  &     39.05  &     765.08  \\
 c10 t0 s q    &     99.93  &      3.22  &     93.62  &      95.59  &     58.42  &     510.38  &     38.10  &     763.64  \\
 \hline
 c10 t0 i w0   & \tb{99.99} &      2.89  &     92.67  &     105.16  &     57.11  &     529.95  &     38.10  &     798.26  \\
 c10 t0 i s w0 &     99.96  &      2.72  & \tb{94.33} & \tb{ 93.99} & \tb{59.47} & \tb{502.30} & \tb{39.05} & \tb{754.46} \\
 \hline
\end{tabular}}\vspace*{1em}
\label{tab:variants}
\end{table}

\begin{table}[p]
\centering
\caption{Relative improvement over \void for selected algorithmic variants. We give the the success rate over the instances unsolved by \void, and
the avg.\ runtime ratio over the commonly solved instances.
}
\scalebox{\tablescale}{\begin{tabular}{|l|r|r|r|r|r|r|r|r|}
 \hline
 \multicolumn{1}{|c|}{variant} & \multicolumn{2}{c|}{Rome} & \multicolumn{2}{c|}{North} & \multicolumn{2}{c|}{Expanders} & \multicolumn{2}{c|}{SteinLib} \\
               & \ce{new [\%]} & \ce{speed-up} & \ce{new [\%]} & \ce{speed-up} & \ce{new [\%]} & \ce{speed-up} & \ce{new [\%]} & \ce{speed-up} \\
 \hline
 c10           &     93.98  & \tb{66.80} &     42.34  &     21.45  &     30.24  &     13.96  &     25.26  & \tb{11.79} \\
 c10 t0 i w0   & \tb{99.92} &     60.85  &     72.07  &     28.59  &     34.27  &     12.68  &     31.58  &      6.79  \\
 c10 t0 i s w0 &     99.75  &     59.58  & \tb{78.38} & \tb{34.03} & \tb{37.90} & \tb{23.21} & \tb{32.63} &      5.42 \\
 \hline
\end{tabular}}
\label{tab:relative}
\end{table}

\begin{table}[p]
\centering
\caption{Average number of cycle variables and average values for maximum cycle length $D$.}
\scalebox{\tablescale}{\begin{tabular}{|lr|r|r|r|r|r|r|r|r|}
 \hline
 variant & & \multicolumn{2}{c|}{Rome} & \multicolumn{2}{c|}{North} & \multicolumn{2}{c|}{Expanders} & \multicolumn{2}{c|}{SteinLib} \\
         \multicolumn{2}{|r|}{min \# var}  & \ce{$D$} & \ce{\#~var} & \ce{$D$} & \ce{\#~var} & \ce{$D$} & \ce{\#~var} & \ce{$D$} & \ce{\#~var} \\
 \hline
 c5  & 500 &  9.51  &    627 &   7.34 &    689 &   5.43 & 2\,075 &   5.80 &    881 \\
 c10 & 1\,000 & 10.51  & 1\,168 &   8.01 & 1\,213 &   5.73 & 2\,816 &   6.64 & 3\,658 \\
 c20 & 2\,000 & 11.51  & 2\,175 &   8.55 & 2\,048 &   6.47 & 7\,774 &   7.09 & 4\,785 \\
 \hline
\end{tabular}}
\label{tab:cycle-lengths}
\end{table}

\subparagraph*{Results.}
Table~\ref{tab:variants} shows the success rates (percentage of instances solved to proven optimality) and average runtime per instance of our algorithmic variants.
For non-solved instances we assume the maximum runtime of 20 minutes---average runtimes are thus comparable only
for algorithms that achieve roughly equal success rates.
We group the variants by the number of used extensions
and highlight variants that dominate their group in bold. The latter informs our choice of which variants to consider in the next group.

The separation of generalized Euler constraints is clearly beneficial only on the North graphs, but even there
its improvements are marginal when compared to the cycle-based approach. The latter works very well in practice, for all instance sets.
In particular (cf.\ Fig.~\ref{fig:plots}), on \emph{Rome} it allows us for the first time to compute the skewness of \emph{all} instances.
Using variant \emph{c10~t0~i~w0}, we are able to solve all but
\texttt{grafo10958.98.lgr} within the 20 minute time frame; this last instance required 103 minutes.
\emph{North} still contains instances too hard to solve exactly (even when increasing runtime to a few days and memory to 32GB).
Nonetheless, we now solve $3 / 4$ of the previously unsolved \emph{North} graphs within our strict limits.
The second group of variants in Table~\ref{tab:variants} demonstrates that all of our extensions of the cycle model,
in particular the pseudo-tree approach, improve upon success rate and runtime on all instance sets when applied to \emph{c10}.
As shown in the lower sections of the table, this does not always apply when comparing models that simultaneously use multiple extensions.

Table~\ref{tab:relative} details the relative improvement for each of the three most promising algorithm configurations over the state-of-the-art \void-model.
We provide the success rate for the instances not solved by \void and
give the average relative speed-up (i.e., the runtime of \void divided by that of variant \emph{X})
over the instances solved by both \void and \emph{X}.
This common set is exactly those solved by \void, except for a single \void-solved \emph{North}-instance 
not solved by \emph{c10}.
On \emph{Rome}, the pure cycle model \emph{c10} without any further extensions achieves the best speed-up; for the seemingly harder other instance sets,
more sophisticated variants are worthwhile. Fig.~\ref{fig:plots} underlines that the success rate of the algorithms is strongly correlated to the instance's skewness.

Table~\ref{tab:cycle-lengths} lists the average number of generated cycle variables and the respective average values for $D$.
We mention that instances with high $D$ values typically generate \emph{few} cycle variables, close to the lower bound.
However, there is a large deviation in the number of generated cycle variables in any fixed instance set:
some graphs contain less than the requested number of cycles whereas others already contain roughly 10\,000 triangles.

\begin{figure}
\centering
\captionsetup[subfigure]{justification=centering}
\newcommand{\mywidth}{.95\textwidth}
\pgfplotscreateplotcyclelist{alg styles}{styleA1,styleB1,styleC1,styleD1}
\doplot[title={Solved Rome graphs by skewness. \emph{c10~t0~i~w0} solves all but 1 skew-$22$-graph within 20min.},
 percent,
 bars,
 width=\mywidth,
 axisArgs = {
  xlabel = skewness,
  ylabel=success rate,
  xtick = {2,4,6,8,10,12,14,16,18,20.5}
 }
]{solved-rome-by-skew}{4}{2,4,6,8,10,12,14,16,18,$\geq\!20$}
\doplot[title={Solved North graphs by best upper bound on skewness.},
 percent,
 bars,
 width=\mywidth,
 axisArgs = {
  xlabel = upper bound on skewness,
  ylabel=success rate,
  xtick = {2,4,6,8,10,12,14,16,18,20,23.5}
 }
]{solved-north-by-skew-ub}{5}{2,4,6,8,10,12,14,16,18,20,$\geq\!23$}
\caption{Detailed success rates for selected algorithmic variants.}\label{fig:plots}
\end{figure}

\section{Conclusion and Open Questions}
For over two decades, the strongest ILP model for MPS has not been improved.
In this paper we presented novel variables and constraints, based on cycles, to extend
this model to finally obtain both a theoretically stronger model, as well as a more efficient
algorithm in practice.
We proved that there is a hierarchy of ever stronger LP-relaxations, induced
 by the maximal considered cycle length, and a rich set of further strengthening cycle-based constraint classes.
For the first time, we are able to compute the skewness of \emph{all} Rome graphs,
 solve 94\% of the North graphs (compared to 74\% by the \void-model), and solve 40\% instead of only 10\% of our SteinLib instances.
 Our extensions also help for the notoriously hard expander graphs.

\smallskip
Several of our proofs show the new constraint class's strength w.r.t.\ a low-$D$ cycle model. We conjecture that most classes remain strengthening
for high $D$, but to prove this, one has to find and argue infinite families of graphs with the LP-properties of our currently hand-crafted
proof graphs. Furthermore, it is natural to ask if and which of the new constraint classes form facets in the (lifted) planar subgraph polytope.

A problem inherent to our approach arises on inputs of non-homogeneous density:
$G$ may have too dense subgraphs to raise $D$ sufficiently,
even when every planar subgraph of $G$ contains large regions consisting of high-degree faces.
Is there a practical way to generalize the cycle-based approach
using an independent maximum cycle length \emph{for each edge}?

\clearpage
\bibliography{main}
\end{document}